\documentclass[twocolumn,astrosym,apj,]{aastex631}
\usepackage{graphicx,amssymb,amsmath}    %% figure package
\usepackage{amsthm,amsfonts,amscd,wasysym,cleveref}
\shorttitle{J0749+2255: Molecular gas feeding two quasars at cosmic noon}
\shortauthors{Ishikawa et al.}
\graphicspath{{./}{figures/}}

\newcommand{\target}{J0749+2255}
\newcommand{\targetfull}{SDSSJ074229.96+225511.7}
\newcommand{\targSW}{J0749+2255-SW}
\newcommand{\targNE}{J0749+2255-NE}

\newcommand{\gaia}{\textit{Gaia}}

\newcommand{\sdss}{\textrm{SDSS}}
\newcommand{\alma}{ALMA}

\newcommand{\ha}{${\rm H\alpha}$}

\newcommand{\hmol}{${\rm H_2}$}
\newcommand{\cii}{[\textrm{C}~\textsc{ii}]}

\newcommand{\pah}{PAH 3.3 $\mu$m}

\newcommand{\coline}{CO$(4-3)$}
\newcommand{\cobase}{CO$(1-0)$}
\newcommand{\colambda}{CO$(4-3)\lambda650\mu$m}

\newcommand{\mum}{$\mu$m}
\newcommand{\kms}{km s$^{-1}$}
\newcommand{\ergs}{erg s$^{-1}$}

\newcommand{\Mbh}{$M_{\textrm{BH}}$}
\newcommand{\Mmol}{$M_{H_2}$}
\newcommand{\Lbol}{$L_{bol}$}

\newcommand{\COcont}{$L_{CO}/L_{\textrm{cont}}$}
\newcommand{\fQratio}{$f_{\rm SW}/f_{\rm NE}$}
\newcommand{\vsigRat}{$v_{50}/\sigma$}

\begin{document}

\title{VODKA: Complex molecular gas dynamics in a kpc-separation $z=2.17$ dual quasar with ALMA}
%\title{Dual or lensed quasar? First look of sub-arc separation $z\sim2$ quasar pair with JWST NIRSpec IFS}
% =================================================================================================================

\newcommand{\jhu}{\rm Department of Physics and Astronomy, Johns Hopkins University, Baltimore, MD 21218, USA}
\newcommand{\mki}{\rm MIT Kavli Institute for Astrophysics and Space Research, Massachusetts Institute of Technology, Cambridge, MA 02139, USA}
\newcommand{\stsci}{\rm Space Telescope Science Institute, 3700 San Martin Drive, Baltimore, MD 21218, USA}
\newcommand{\ias}{\rm Institute for Advanced Study, Princeton University, Princeton, NJ 08544, USA}
\newcommand{\uiuc}{\rm Department of Astronomy, University of Illinois at Urbana-Champaign, Urbana, IL 61801, USA}
\newcommand{\heidelberg}{\rm Zentrum für Astronomie der Universität Heidelberg, Astronomisches Rechen-Institut, Mönchhofstr 12-14, D-69120 Heidelberg, Germany}

% ------------------------------------------------------------------------------------------------------

\correspondingauthor{Yuzo Ishikawa}
%\email{yishika2@jhu.edu}
\email{yishika2@mit.edu}

\author[0000-0001-7572-5231]{Yuzo Ishikawa}
\affiliation{\jhu}
\affiliation{\mki}

\author[0000-0001-6100-6869]{Nadia L. Zakamska}
\affiliation{\jhu}

\author[0000-0002-9932-1298]{Yu-Ching Chen}
\affiliation{\jhu}

\author[0000-0002-0710-3729]{Andrey Vayner}
\affiliation{\jhu}
\affiliation{\rm IPAC, California Institute of Technology, 1200 E. California Boulevard, Pasadena, CA 91125, USA}

\author[0000-0003-1659-7035]{Yue Shen}
\affiliation{\uiuc}

\author[0000-0003-0049-5210]{Xin Liu}
\affiliation{\uiuc}
\affiliation{\rm National Center for Supercomputing Applications, University of Illinois at Urbana-Champaign, Urbana, IL 61801, USA}
\affiliation{\rm Center for Artificial Intelligence Innovation, University of Illinois at Urbana-Champaign, 1205 West Clark Street, Urbana, IL 61801, USA}

\author[0000-0003-4250-4437]{Hsiang-Chih Hwang}
\affiliation{\rm School of Natural Sciences, Institute for Advanced Study, 1 Einstein Drive, Princeton, NJ 08540, USA}

\author[0000-0002-4419-8325]{Swetha Sankar}
\affiliation{\jhu}

\author[0000-0001-7681-9213]{Arran C. Gross}
\affiliation{\uiuc}

\iffalse
\author[0000-0002-1608-7564]{David S. N. Rupke}
\affiliation{\rm Department of Physics, Rhodes College, 2000 N. Parkway, Memphis, TN 38112, USA}
\affiliation{\heidelberg}

\author[0000-0002-3158-6820]{Sylvain Veilleux}
\affiliation{\rm Department of Astronomy and Joint Space-Science Institute, University of Maryland, College Park, MD 20742, USA}

\author[0000-0003-2212-6045]{Dominika Wylezalek}
\affiliation{\heidelberg}

\author[0000-0001-7681-9213]{Arran C. Gross}
\affiliation{\uiuc}

\author{Nadiia Diachenko}
\affiliation{\jhu}

\fi
%\author{\red{VODKA Team}}

% =================================================================================================================

\begin{abstract}
In galaxy mergers, dual quasars — two actively accreting supermassive black holes (SMBHs) — provide a unique opportunity to study the interplay between galaxy dynamics and quasar activity. However, very little is known about their molecular gas, which fuels star formation and quasar activity. In this study, we map the kinematics of the cold molecular gas in \target, a 3.8 kpc separation dual quasar at $z=2.17$ using the Atacama Large Millimeter Array (\alma) Band 4. We detect \colambda, which shows remarkably complex morphological and kinematic structures. While the integrated CO map suggested a lens-like ring, this feature disappears with kinematic decomposition. The kinematic analysis with ALMA resolves the ambiguities introduced by previous observations, further supporting the dual quasar interpretation of \target. We find two kinematically distinct molecular gas components: spatially extended, yet dynamically complex slow-moving gas ($\textrm{FWHM}\sim 130\ \textrm{km s}^{-1}$), and a compact, blueshifted, fast-moving, turbulent gas ($\textrm{FWHM}\sim300\ \textrm{km s}^{-1}$). The disturbed kinematics, likely driven by the merger, show hints of rotation but no molecular outflows, suggesting circumnuclear flows. We estimate a large molecular gas reservoir ($M_{\textrm{H}_2}\sim10^{10}M_{\odot}$), yet the starburst activity appears to exceed the available fuel. We detect an extended continuum in excess at rest-frame 455 GHz. The kinematic complexity of CO implicates the connection of mergers on the starburst and quasar activity in \target, yet whether \target\ represents the dual quasar population remains unclear. Targeted kinematic studies of larger dual quasar samples will be essential to disentangling the nature of dual quasars.
\end{abstract}

%\keywords{galaxies: active –- quasars: general -- quasars: supermassive black holes }


\section{Introduction} \label{sec:intro}
Over the past several decades, there has been increasing evidence for the co-evolution of galaxies and the supermassive black holes (SMBHs) they harbor \citep{Silk1998, DiMatteo2005}. Following a merger of two galaxies, the two central SMBHs may spiral into the center through dynamical friction and interaction with the gas and stars to form a bound binary  \citep[e.g.][]{Begelman1980, Milosavljevic2001, Blaes2002, Yu2002a}. When two SMBHs in merging galaxies are actively accreting during inspiral, they can be observed as a dual quasar (active galactic nuclei; AGN). Although there is some evidence that quasar activity may be enhanced in dual SMBHs due to gas flows during a merger \citep[e.g.][]{Begelman1980, Hopkins2009} the mechanisms for quasar triggering remain poorly understood and hotly debated \citep[e.g.][]{Mechtley2016}. Gas infall may be triggered by merger-induced tidal torques or even secular accretion. Since dual quasars are found in galaxy mergers, studying dual quasars may help unravel the poorly understood connection between SMBH fueling, quasar triggering, and galactic dynamics. 

Whatever the fueling mechanism, quasar activity requires an abundant supply of cold, dense gas transported to the circumnuclear regions \citep{Heckman2014}. Dense molecular gas reservoirs provide fuel for both star formation \citep{Carilli2013} and quasar activity. This means that the fate of the molecular gas phase is a key link in understanding the relationship between quasars and their host galaxies. Typically, the second most abundant molecule, carbon monoxide (CO), is used as a tracer of the molecular gas, which is dominated by molecular hydrogen (\hmol). Unlike \hmol, CO has a much lower excitation energy, requiring gas temperatures above $\sim5$ K, which makes it easier to observe \citep{Carilli2013}. Observations of the cold molecular gas phase at high sensitivity and angular resolution have been enabled by observatories such as ALMA. 

Over the past several years, extensive studies have explored the molecular properties of quasar host galaxies. Past studies of quasar hosts have revealed diverse and, at times, contradictory molecular gas properties. Kinematic measurements revealed compact hosts with rotating disks and irregular morphologies with disturbed kinematics \citep[e.g.][]{DiazSantos2016, Banerji2017, Brusa2018, Feruglio2018, Bischetti2021}. Intense star formation rates (SFRs), up to several $\gtrsim 1,000\ M_{\odot}\textrm{ yr}^{-1}$, have been observed in some host galaxies \citep{Maiolino2012, Duras2017, FanL2019, Nguyen2020}. Some studies find that quasar hosts exhibit higher star formation efficiency than galaxies without a quasar \citep[e.g.][]{Bischetti2017}, but others did not find any significant differences \citep[e.g.][]{Kirkpatrick2019, Valentino2021}. Quasar hosts have been found to be CO/molecular gas depleted \citep[e.g.][]{Kakkad2017, Brusa2018, Bischetti2021, Vayner2021a, Bertola2024}, raising the question whether quasar host galaxies lie on or off the ``main sequence'' of star formation \citep[e.g.][]{Xie2021}. It is unclear if these variations are due to quasar feedback in the form of molecular outflows or due to galaxy interactions \citep[e.g.][]{Kakkad2017, Trakhtenbrot2017, Bischetti2018, Brusa2018, FanL2019, HerreraCamus2019, Vayner2021a}. 

Despite progress in studying the molecular gas properties of quasar host galaxies, many questions remain unanswered — especially for dual quasars, whose overall properties are still poorly understood. The challenges stem from difficulties in identifying reliable candidates and disentangling the quasar and host galaxy light. However, recent advances in spatially resolved imaging and spectroscopy have begun to shed light on their rest-frame ultraviolet and optical properties. Some reveal complex merger activity \citep{Treister2018, Koss2018, Tubin2021}, some reveal intense star formation with disk-dominated host galaxies \citep{ChenYC2023a, ChenYC2024, Ishikawa2024}, and some reside in complex environments \citep{Perna2023}. Limited studies of the molecular gas have detected CO and \cii$158\ \mu\textrm{m}$ emission \citep[e.g.][]{Treister2018, Koss2023, Tang2024}. Some studies inferred large molecular gas reservoirs exceeding $10^9\ M_{\odot}$ \citep{Treister2018, Tang2024}. Complex molecular gas dynamics, including velocity gradients, tidal tails, and even gas bridges that connect the two quasars have been detected \citep{Izumi2024}, possibly linked to the ongoing merger activity \citep{Treister2018, Koss2023}. Quasar feedback has been suggested in some dual quasar systems \citep{Tang2024, Izumi2024}, yet their prevalence remains unclear. 

In this paper, we obtain and analyze spatially resolved \alma\ observations of the molecular gas traced with CO in \targetfull\ (\target, henceforth), a dual quasar at $z=2.17$. Although the CO($1-0$) transition is ideal for tracing the total molecular gas, we choose the \colambda\ transition, the best observable line during Cycle 9. This paper aims to probe the impact of galaxy mergers on the host galaxies and their central SMBHs.  In Section \ref{sec:dataRedux} we summarize the known properties of \target\ and outline the new \alma\ observation and data reduction. Throughout this paper, we refer to both dual AGNs and dual quasars as ``dual quasars'' for simplicity. In Section \ref{sec:specAnaly} we present the spectral analyses. We discuss the interpretation of the data in Section \ref{sec:discuss}, and conclude in Section \ref{sec:concl}. We adopt the $\Lambda$CDM cosmology with $h = 0.7$, $\Omega_M = 0.3$, and $\Omega_{\Lambda} = 0.7$. 

\section{Observations and data reduction }\label{sec:dataRedux}

\subsection{Summary of J0749+2255}
In the last decade, techniques have been significantly refined to search for close quasar pairs systematically. The advent of large-scale imaging surveys such as Subaru/HSC \citep{HSC2018} and \gaia\ \citep{Gaia2016} allows for a systematic search for close quasar pairs. In particular, two methods using \gaia\ have pioneered the discovery of sub-arcsec pairs: Varstrometry for off-nucleus and dual subkiloparsec AGN (VODKA) and Gaia Multi Peak (GMP), which capitalize on the superb precision astrometry and excellent point-spread function of \gaia\ \citep{Shen2019a, Hwang2020, Mannucci2022}. The method of interest in this paper is VODKA, which searches for light centroid jitters caused by the asynchronous stochastic variability of unresolved quasar pairs \citep{Shen2019a, Hwang2020}. These \gaia-based methods have been successful in identifying dual quasar candidates with $<1''$ or kpc-scale separations. Follow-up studies have confirmed some of these candidates as bona fide dual quasars, lensed quasars, and star-quasar superpositions \citep[e.g.][]{ChenYC2022, Ciurlo2023, Scialpi2024}.

\target\ is a VODKA-selected dual quasar at $z=2.17$. It was initially spectroscopically identified as a Type 1 quasar by the Sloan Digital Sky Survey (\sdss; \citealt{Schneider2010}). Follow-up Hubble Space Telescope (HST) and VLA-A (6 GHz and 15 GHz) observations revealed two point-like cores separated by $\sim0.5''$, corresponding to a physical separation of $\sim3.8\textrm{ kpc}$, \citep{Shen2021, ChenYC2022, ChenN2023a}. The brighter quasar is located to the southwest (\targSW, henceforth) and the fainter quasar is located to the northeast (\targNE, henceforth). \cite{ChenYC2023a} outlined the first series of follow-up observations to confirm the nature of the \target. Spatially-resolved slit-spectroscopy with Gemini/GMOS+GNIRS revealed two distinct quasar spectra. Furthermore, detailed point-spread-function (PSF) modeling of the HST images hinted at extended tidal features that are likely part of an ongoing merger. 

\begin{table*}
\centering
\caption{Summary of \alma\ observations. The observed dates include calibration and on-source pointings.} 
\label{tab:obsv} 
\begin{tabular}{lcc}
    \hline
    Parameter  & Value  \\
    \hline
    Observed Dates  & 2023-04-22, 2023-05-13, 2023-05-14 \\
    Continuum Central Frequency & 143.538 GHz \\
    Continuum Beam & $0.292'' \times 0.252''$ \\
    Continuum Sensitivity &  $0.013$ mJy/beam over 2.95 MHz \\
    Line Central Frequency &  145.434 GHz   \\
    Line Beam &  $0.296'' \times 0.254''$ \\
    Line Sensitivity &  $0.094$ mJy/beam over 15.6275 MHz \\
    Channel Width & 32 \kms\ per channel \\
    \hline
\end{tabular}
\end{table*}

Recently, spatially resolved spectroscopy with the James Webb Space Telescope (JWST) detected extended ionized optical and molecular infrared gas associated with the host galaxy using NIRSpec \citep{Ishikawa2024} and MIRI \citep{ChenYC2024}. The JWST observations revealed three key results. First, \target\ is hosted by a powerful starburst galaxy with an SFR exceeding $1,000\ M_{\odot}\textrm{ yr}^{-1}$, based on the ionized \ha\ \citep{Ishikawa2024} and molecular \pah\ \citep{ChenYC2024} emission lines. The optical line ratio diagnostics indicate that these star-forming regions are concentrated close to the two quasars \citep{Ishikawa2024}. Second, the spatially-resolved kinematic analysis of the emission lines suggests the presence of a large rotating gas with a symmetric blue-/redshifted velocity gradient spanning 10 kpc in diameter \citep{Ishikawa2024}. This suggests \target\ is a dual quasar system either hosted by a single galaxy or a merger with complex kinematics. 

Third, an unexpected result is that the two quasars have remarkably similar, yet subtly different, spectra and black hole properties, \Mbh\ and luminosity \citep{Ishikawa2024}. The natural explanation of the apparent quasar similarity is gravitational lensing instead of a physical dual quasar pair. Although lens modeling of the HST images \citep{ChenYC2023a} and JWST kinematic analysis of the two quasars and the extended ionized gas \citep{Ishikawa2024} argued against the lensing scenario, JWST's improved aperture spectroscopy resurfaced the lensing hypothesis. Interestingly, this strange scenario of similar dual quasars hosted by a single galaxy is predicted by numerical simulations \citep[e.g.][]{Dadiani2024}. Thus, distinguishing between a dual quasar and a lensed double quasar is not clear-cut. In fact, \cite{Gross2023} stresses that exhaustive multi-wavelength analysis is necessary to resolve the ambiguity of close-separation pairs.

Under the lensing hypothesis, reconciling the JWST observations of the extended host galaxy gas with a lensing model remains a key challenge. Specifically, how to split the quasar images while maintaining the kinematics of the large-scale gas disk. To further elucidate the nature of \target, we use \alma\ data to study the cold molecular gas within the framework of two competing interpretations: dual quasar vs.~lensed quasar. %Perhaps the biggest advantage of ALMA is its high sensitivity and angular resolution.

\subsection{ALMA: Observation and data reduction}
The leading goal of the \alma\ observation was to detect the \coline\ and the continuum at the same angular resolution as HST and JWST/NIRSpec. \target\ was observed with ALMA Band 4 in Cycle 9 (2022.1.01427.S; PI: Ishikawa) over three blocks with the C-6 configuration for an effective exposure time of 8,164.74 seconds. This corresponds to an angular resolution of $\sim 0.2''$ $(1.66\ \textrm{kpc})$. %and a maximum recoverable scale of \red{$3.0''-4.1''$ $(25-34\ \textrm{kpc})$}. 
The spectral windows were tuned to the redshifted frequency of \coline\ at 145.434 GHz and the nearby continuum at 143.538 GHz. 

We reduce the \alma\ data using the Common Astronomy Software Applications (CASA) version 6.4.1.12 following the procedures in \cite{Vayner2021a}. The final datacube has a spatial resolution of $0.05''$ per pixel, which corresponds to a plate-scale of $\sim 0.5\ \textrm{kpc/pixel}$. 

\section{Spectral Analysis}\label{sec:specAnaly}

\subsection{Continuum emission}\label{subsec:specCONT}
\begin{figure}%[!hbt]
	 \begin{center}
    \includegraphics[width=\columnwidth,trim={0.1cm 0 0.8cm 0},clip]{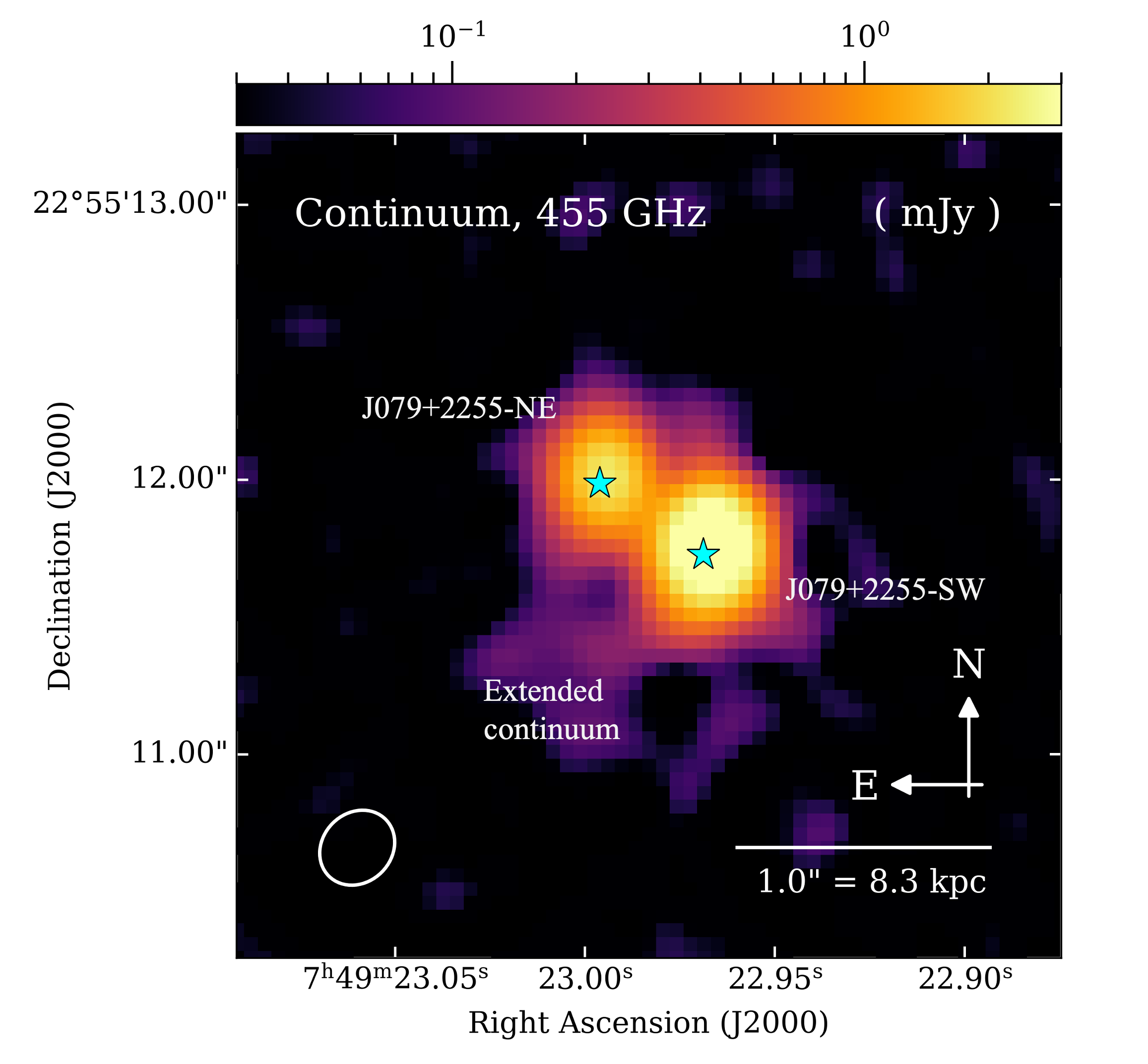}
	 \end{center}
	 \caption{The continuum image centered at the observed frequency 143.538 GHz. The cyan stars indicate the known quasar positions. There are two components to the continuum: the two bright cores corresponding to \targSW\ and \targNE\ and a faint, extended component stretching towards the southeast. We also show the beam size.} 
	 \label{fig:cont} 
\end{figure}

In Figure \ref{fig:cont} we show the continuum emission centered on the rest-frame frequency of 455.02 GHz (658 \mum), observed at 143.538 GHz (2088 \mum) over a bandwidth of 2 GHz. We see two bright clumps that are co-spatial with the known positions of the two quasars and a fainter, extended component stretching asymmetrically to the southwest relative to the two quasars. 

\begin{table*}
    \centering
    \caption{Continuum at 143.538 GHz observed frequency. We take $r=1''$ apertures to calculate the total continuum, which includes the two quasars and the extended component. We also extract the continuum fluxes co-spatial with the two quasars with $r=0.15''$, which is roughly the half-light radius. } 
    \label{tab:cont} 
    %\resizebox{\columnwidth}{!}{
    \begin{tabular}{lccc}
        \hline
         & Total Continuum & \targSW\ & \targNE \\
        %Aperture radius & ($1.0''$)  & ($0.20''$) & ($0.15''$) \\
        \hline
        Flux (Jy) & $0.016 \pm 0.002$ & $0.0049 \pm 0.0005$ & $0.0020 \pm 0.0002$ \\
        $\nu L_{\nu}$ (\ergs)  & $(2.52 \pm 0.35)\times10^{45}$ & $(7.9 \pm 0.9)\times10^{44}$ & $(3.2 \pm 0.5)\times10^{44}$ \\
        $\nu L_{\nu}$ ($L_{\odot}$)   & $(6.6 \pm 0.9)\times10^{11}$ & $(2.1 \pm 0.2)\times10^{10}$ & $(8.3 \pm 1.2)\times10^{10}$ \\
        \hline
    \end{tabular}
    %}
\end{table*}

We also perform aperture photometry of the continuum centered on each quasar nuclei with aperture radii of $0.15''$ for both \targSW\ and \targNE. We compute the monochromatic luminosity co-spatial with each quasars, $\nu L_{\nu}$, of $(7.9\pm0.9)\times10^{44}\ \textrm{erg s}^{-1}$ and $(3.2\pm0.5)\times10^{44}\ \textrm{erg s}^{-1}$, respectively. The flux ratio of the two quasars' continuum emission (\fQratio) is $\sim2.5$, which differs from the previously measured \fQratio\ at rest-frame optical \citep{ChenYC2023a, Ishikawa2024}.

The extended continuum component stretches to $1''$ or $8\ \textrm{kpc}$ from the quasars. We integrate over the $r=1''$ aperture centered at the midpoint of the two quasars to calculate the total continuum flux of $16 \pm 2\ \textrm{mJy}$, which corresponds to a monochromatic luminosity of $\nu L_{\nu}=(2.52\pm0.35)\times10^{45}\ \textrm{erg s}^{-1}$ or $(6.6\pm0.9)\times10^{11}\ L_{\odot}$. To calculate the uncertainty, we compute the root-mean-square value across a large annular aperture centered on \target, from $r=1.75''$ to $20''$. Then we also assume a conservative 10\% uncertainty in the ALMA flux calibration. We summarize the measured continuum fluxes and calculated luminosities in Table \ref{tab:cont}.

\begin{figure*}
    \begin{center}
    \begin{tabular}{cc}
    \includegraphics[width=0.45\textwidth,trim={0.1cm 0 1.25cm 0},clip]{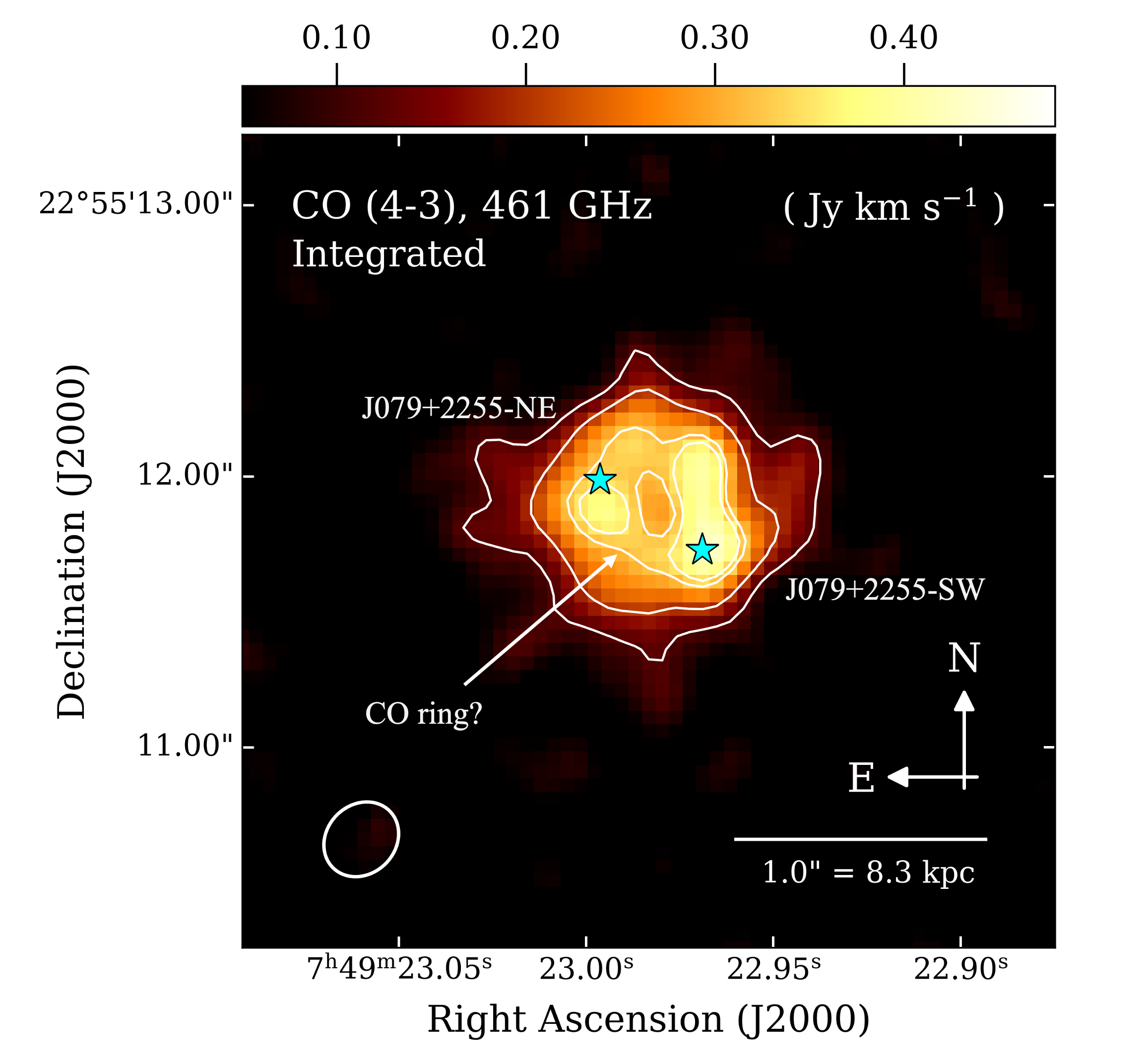} &
    \includegraphics[width=0.50\textwidth]{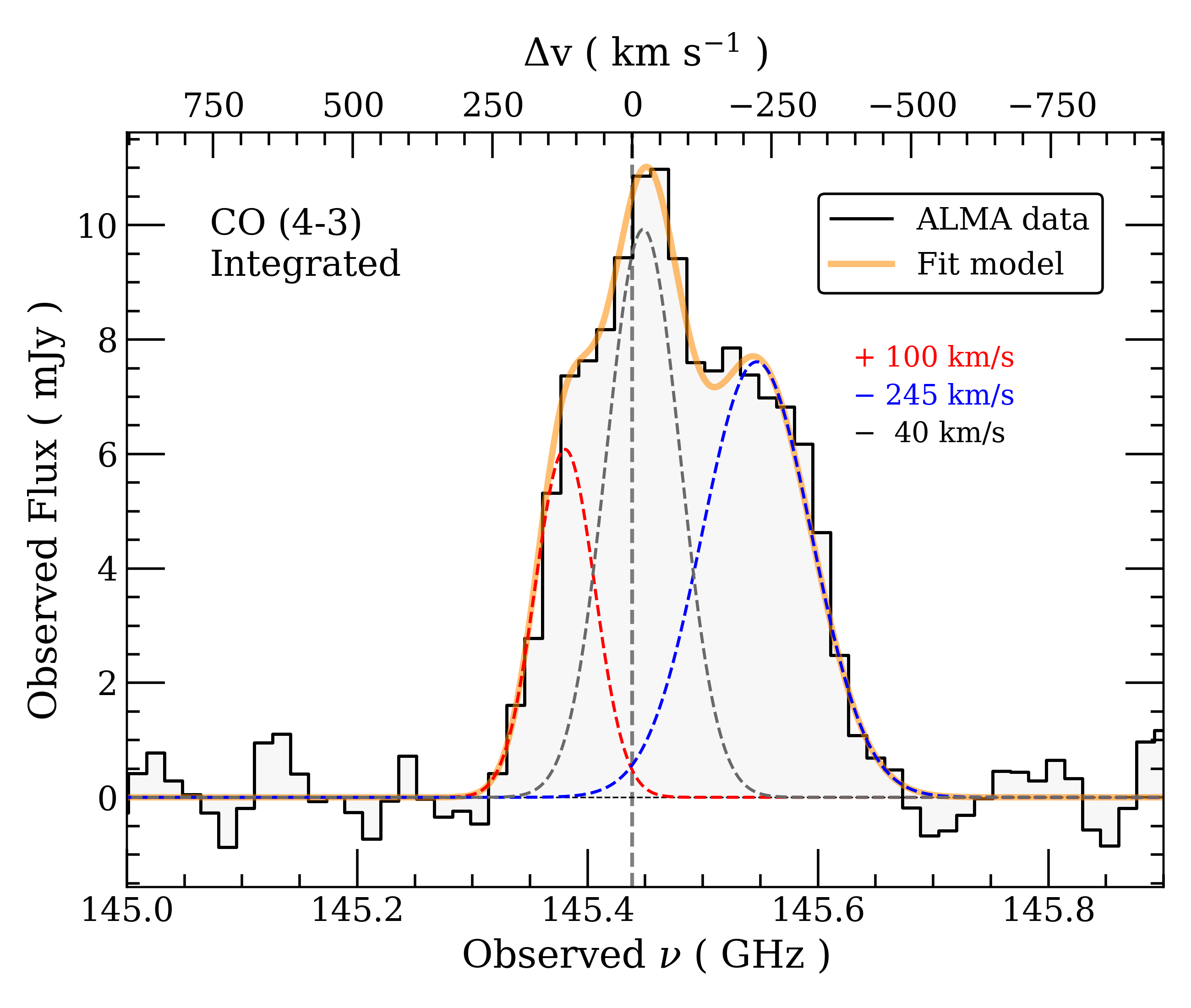} 
    \end{tabular}
     \end{center}
    \caption{(Left) The continuum-subtracted, velocity-integrated \coline\ line intensity over $\sim0.39\ \textrm{GHz}$ ($\sim800\ \textrm{km s}^{-1}$). The integrated intensity map is shown in logscale. The cyan stars indicate the known quasar positions. The CO-emitting gas has an extended component out a radius of $\sim0.6''$ and a bright, compact CO ring concentrated within a radius of $\sim0.3''$. The white contour lines are set at 0.12, 0.20, 0.32, and 0.35 $\textrm{Jy km s}^{-1}$ to highlight the CO ring structure. The brightest CO clumps are offset from the known quasar positions. We indicate the beam size in the bottom left corner. (Right) The integrated \coline\ spectrum, extracted from the $1''$ aperture radius, in the observed frame with the corresponding velocities relative to the rest frame frequency is shown. The integrated spectrum is fit with three Gaussian components, which indicates distinct velocity components: a bright peak centered near $z_{sys}$, a faint redshifted component with $\Delta v=100\ \textrm{km s}^{-1}$, and a broad blueshifted component $\Delta v=-245\ \textrm{km s}^{-1}$.}
    \label{fig:almaCO43} 
\end{figure*}

\subsection{Morphological Analysis of Integrated CO(4-3)}
\label{subsec:spectanaly}

\begin{figure*}
    \begin{center} 
    \includegraphics[trim={0.35cm 0.35cm 0.35cm 0.35cm},clip,width=0.8\textwidth]{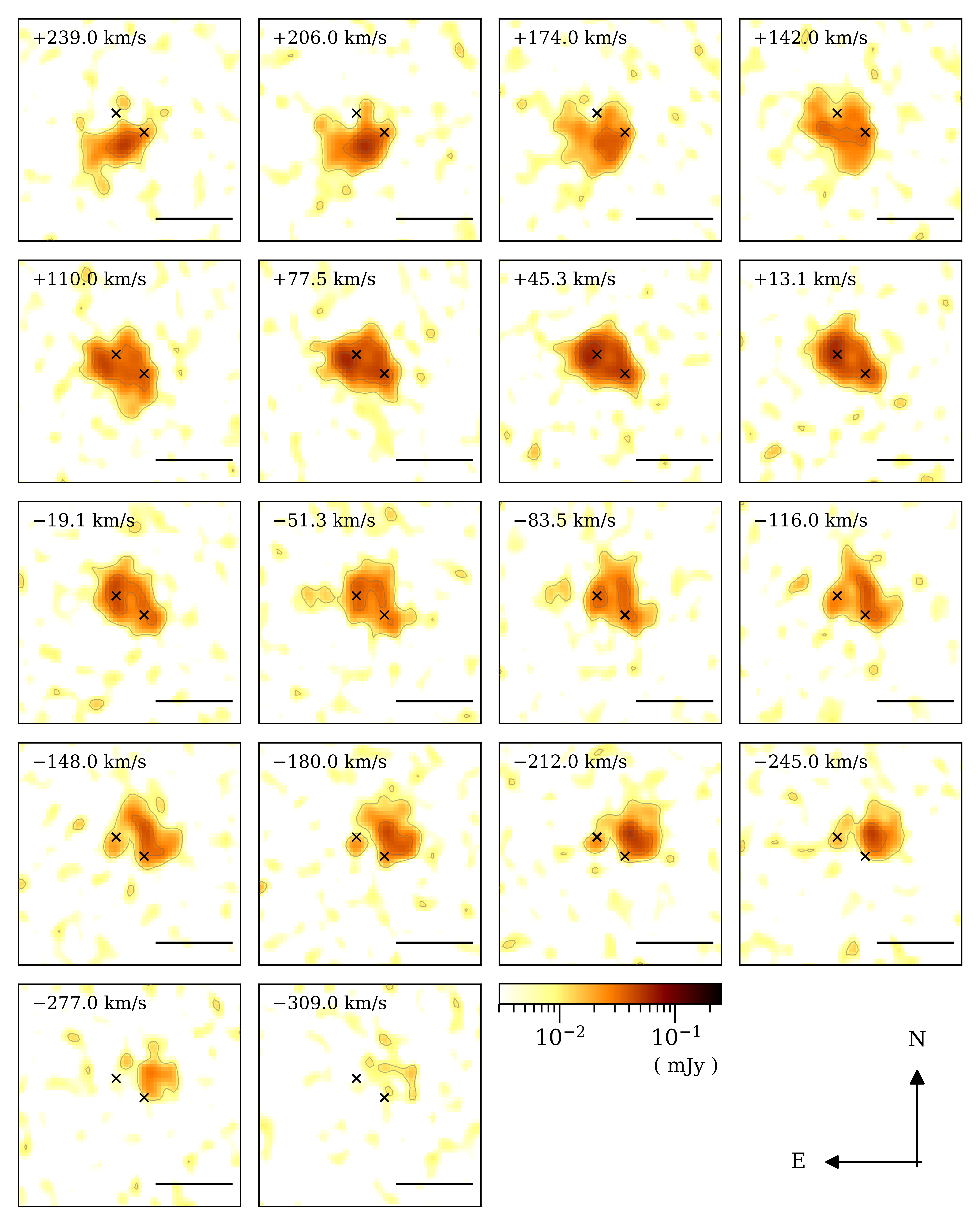}
     \end{center}
    \caption{The velocity channel map of the observed \coline. We calculate the velocity offset with respect to the rest-frame frequency, assuming $z=2.169$. Each slice corresponds to the channel width of $\Delta v \sim 32\ \textrm{km s}^{-1}$. We see different velocity-dependent structures. A clumpy, CO ring is notable around $\Delta v\sim0\ \textrm{km s}^{-1}$. Extended blue-/red-shifted features are prominent at $\Delta v\sim\pm200\ \textrm{km s}^{-1}$ that appear to be associated with \targSW. Each figure has the same scale size as Figures \ref{fig:almaCO43}. The faint gray contour traces peak channel fluxes at $0.01$, $0.03$, and $0.05\ \textrm{mJy}$. All panels share the same log color scale. The solid black scale line indicates $1''$, and the cross-marks indicate the two quasars.}
    \label{fig:coChanMAP} 
\end{figure*}

In Figure \ref{fig:almaCO43}, we show the integrated intensity map and the integrated spectrum of \coline. The CO map is integrated over $\Delta v\approx800\ \textrm{km s}^{-1}$. The spectrum is extracted from a circular aperture with a radius $r=1''$ or $8\ \textrm{kpc}$ centered at the midpoint of the two quasars. The integrated CO map and spectrum reveals morphologically and kinematically complex features. The integrated flux map shows a bright, clumpy, and compact ring-like structure (``CO ring'' henceforth) at the center within a $0.3''$ (2.48 kpc) radius, surrounded by a fainter, extended component reaching out to $0.6''$ (4.96 kpc). Assuming this CO ring is an ellipse, the angular size of the semi-major and semi-minor axes are $0.3''\times0.2''$ ($1.656\  \textrm{kpc}\times 2.483\ \textrm{kpc}$) oriented at $\sim 45^{\circ}$ east of north. The CO ring also appears to connect the two quasars.

\begin{table}
\centering
\caption{We summarize the integrated \coline\ emission properties (flux, luminosity, and line width). The $w_{80}$ corresponds to the 80th-percentile line width of a non-parameteric three Gaussian component fit of the integrated spectrum. The errors represent the $1\sigma$ uncertainty.} 
\label{tab:CO43} 
\begin{tabular}{ll}
\hline
Parameter & Value \\
\hline
$S_{CO(4-3)}\Delta v$ & $1.36 \pm 0.1\ \textrm{Jy km s}^{-1}$ \\
$L'_{CO(4-3)}$        &  $(1.94 \pm 0.14) \times 10^{10}\ \textrm{K km s}^{-1}\ \textrm{pc}^2$ \\
$L_{CO(4-3)}$         & $(6.07 \pm 0.9) \times 10^7\ L_{\odot}$ \\
$w_{80}$              & $365\pm 5\ \textrm{km s}^{-1}$ \\
\hline
\end{tabular}
\end{table}

The integrated spectrum reveals a broad and complex emission line profile. We fit the integrated spectrum and find that the best model requires at least three Gaussian components - a bright component blueshifted at $-40\ \textrm{km s}^{-1}$, a blueshifted component at $-245\ \textrm{km s}^{-1}$, and a fainter redshifted component at $v=100\ \textrm{km s}^{-1}$ (Figure \ref{fig:almaCO43}). These line components have similar velocity dispersions ($\sim30\ \textrm{km s}^{-1}$, $\sim20\ \textrm{km s}^{-1}$, and $\sim15\ \textrm{km s}^{-1}$, respectively). The fit slightly improves with an additional Gaussian component, but the fitting parameters become degenerate with one another. To quantify the effective, non-parametric line width, we quote the effective 80-percentile line width ($w_{80}$; \citealt{Zakamska2014}) of the integrated CO spectrum. First, we construct the cumulative flux distribution from the best-fit three Gaussian profiles. Then we determine the 10th and 90th percentile velocities of the total flux, $v_{10}$ and $v_{90}$, where $w_{80}$ is the difference $w_{80} = v_{90} - v_{10}$. For a single Gaussian, $w_{80}=2.563\ \sigma=1.088\ \textrm{FWHM}$. We find widths reaching $w_{80} \approx 350\ \textrm{km s}^{-1}$, whereas the cold gas in typical massive galaxies has line widths $w_{80}< 200\ \textrm{km s}^{-1}$ \citep[e.g.][]{He2023, Rizzo2023} after converting from the published full-width-at-half-max (FWHM) values. Although luminous quasars are known to have broader molecular line widths \citep[e.g.][]{Bischetti2021}, these are typically associated with fast-moving outflows. These fit results suggest that the integrated CO spectrum captures a kinematically complex emission, which makes it difficult to assess the nature of the target with the integrated spectrum alone. We explore the spatially resolved kinematic analysis in Section \ref{subsec:fullspec}.

\begin{figure*}
    \begin{center} 
    \includegraphics[width=0.95\textwidth]{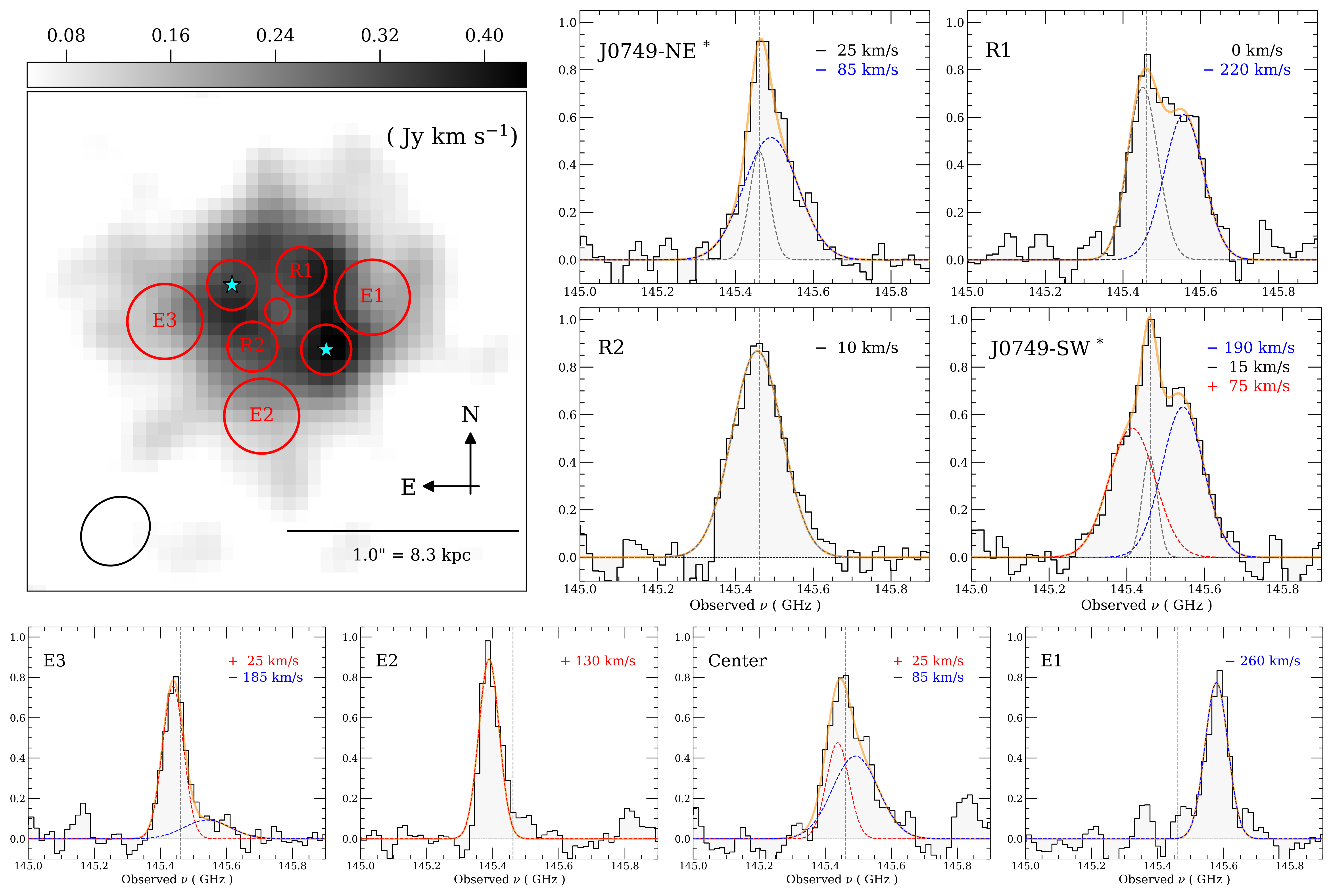}
     \end{center}
    \caption{(Top left) Zoom-in of the integrated CO intensity from Figure \ref{fig:almaCO43} shown in grayscale, labeled with seven red circular apertures for spectral extraction. (1) the apertures co-spatial with the quasars \targSW\ and \targNE, which are indicated with cyan stars; (2) two additional apertures along the inner CO ring, labeled R1 and R2; (3) an aperture at the midpoint of the two quasars and the center of the CO ring; and (4) three apertures from the extended blueshifted and redshifted regions, labeled E1, E2, and E3. The red circle sizes reflect the aperture radii. Apertures SW/NE/R1/R2 were taken at $r=0.1''$, apertures E1/E2/E3 were taken at $r=0.15''$, and the unmarked Center aperture was taken at $r=0.05''$ to avoid overlap. (Top right and bottom row) We show the corresponding aperture spectra. The spectra of \targSW, \targNE, R1, and R2 are shown to match the relative spatial alignment; E1, E2, and E3 are shown on the bottom row. For each aperture spectra, we show the corresponding best spectral fits. The flux values shown are normalized by the peak flux of \targSW\ to compare the relative fluxes and spectral shapes. We can see significant spectral variations between \targSW\ and \targNE. }
    \label{fig:aperspect} 
\end{figure*}

Next, we calculate the integrated \coline\ luminosity, $L'_{CO(4-3)}$. Based on the line fits, we measure the total line-integrated flux $S_{CO}\Delta v= 1.36\pm0.10\ \textrm{Jy km s}^{-1}$ and convert it to $L'_{CO(4-3)}$ in units of $\textrm{K km s}^{-1} \textrm{ pc}^2$ \citep{Solomon2005, Carilli2013}:
\begin{equation}
        L'_{CO(4-3)} = 3.25\times10^7 S_{CO}\Delta v \frac{D_L^2}{(1+z)^3\nu_{obs}^2} \textrm{K km s}^{-1} \textrm{ pc}^2,
    \label{eq:colum}
\end{equation}
where $\nu_{obs}$ is the observed \coline\ frequency in GHz, $D_L$ is the luminosity distance in Mpc, and $z$ is the redshift of \target. We obtain $L'_{CO(4-3)}=(1.94 \pm 0.14)\times10^{10}\  \textrm{K km s}^{-1} \textrm{ pc}^2$, which corresponds to $L_{CO(4-3)}=(6.1\pm 0.9)\times10^7\ L_{\odot}$. The flux uncertainty is determined from the line fit errors and is propagated through the $L'_{CO(4-3)}$ calculation. Table \ref{tab:CO43} summarizes the integrated line values. %We explore the physical interpretation of the \coline\ emission, including the molecular mass estimates, later in Section \ref{sec:DISC:co}.

\begin{figure*}
    \begin{center} 
    \includegraphics[width=0.89\textwidth,trim={0.2cm 0 0.2cm 0cm},clip]{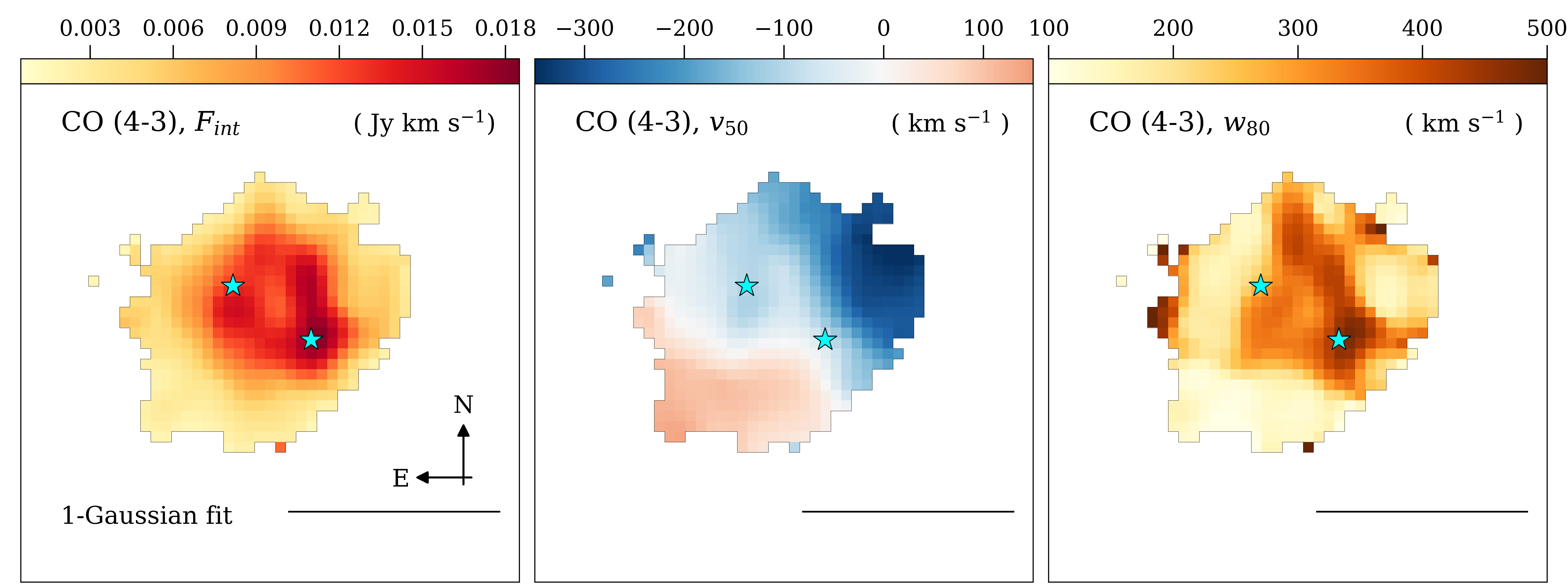}
     \end{center}
    \caption{\coline\ moment maps assuming single-component Gaussian emission line fits for each spaxel. (Left) $F_{int}$ integrated line intensity; (Center) $v_{50}$ velocity shift, and (Right) $w_{80}$ line width. The CO $v_{50}$ distribution roughly matches the kinematics of the ionized gas as seen with JWST \citep{Ishikawa2024}, although with some differences. The large $w_{80}$ values roughly trace the CO ring seen in Figure \ref{fig:almaCO43}. The cyan stars indicate the known quasar positions. The solid lack scale line indicates $1''$.}
    \label{fig:n1fits} 
\end{figure*}

\begin{figure*}
    \begin{center} 
     \includegraphics[width=0.92\textwidth,trim={0.5cm 0 0.1cm 1.5cm},clip]{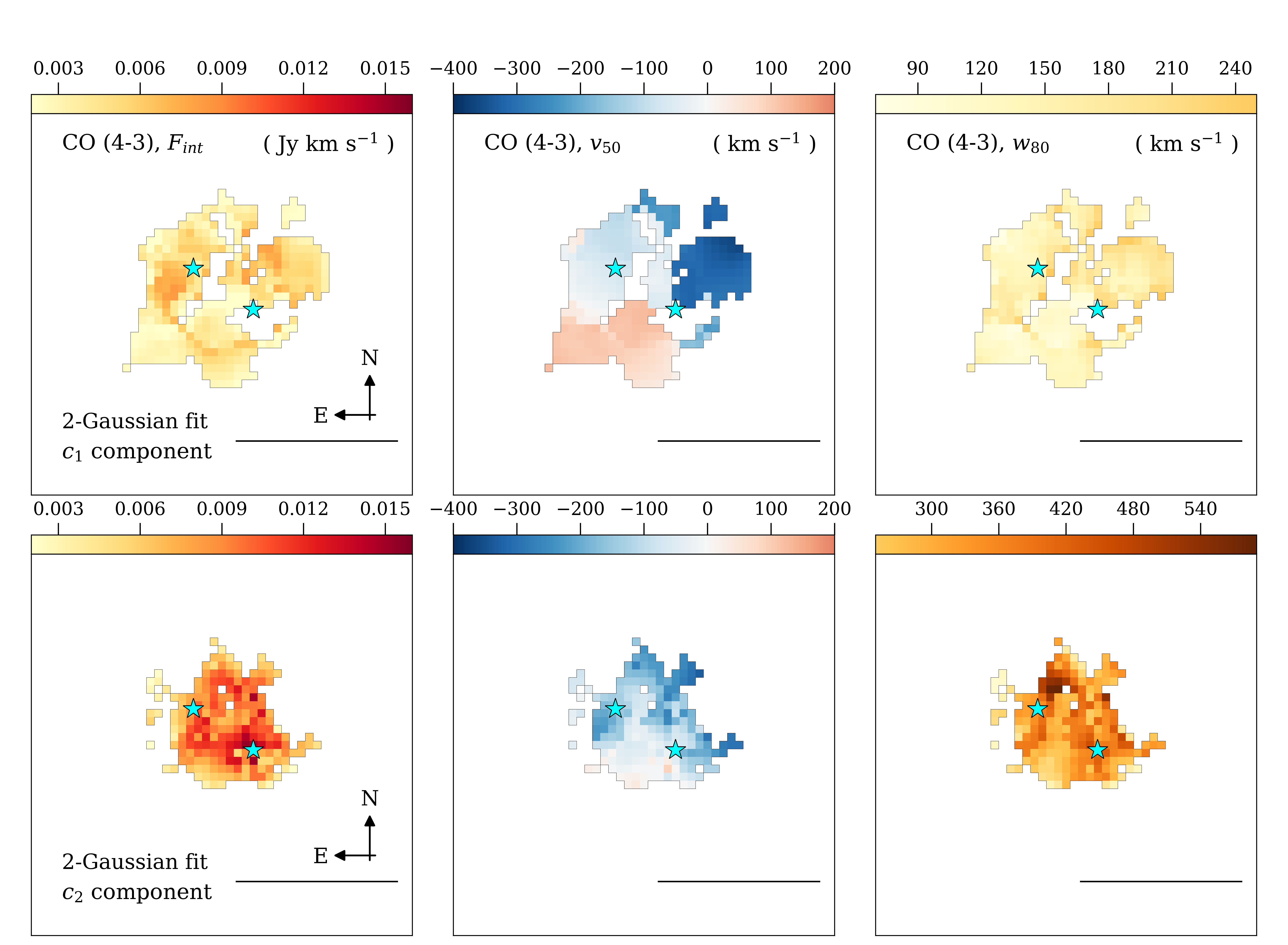}
     \end{center}
    \caption{\coline\ moment maps for two-component ($c_1$, $c_2$) Gaussian line fits. We sort the components by line widths, in which $c_1$ has $\sigma<100\ \textrm{km s}^{-1}$ and $c_2$ has $\sigma>100\ \textrm{km s}^{-1}$. (Top left/center/right) $F_{int}$, $v_{50}$, and $w_{80}$ moment maps for the $c_1$ component. (Bottom left/center/right) The moment maps for the $c_2$ component. The spectral decomposition eliminates the large $w_{80}$ CO ring in Figure \ref{fig:n1fits}. The cyan stars indicate the known quasar positions. The solid lack scale line indicates $1''$.}
    \label{fig:n2fits} 
\end{figure*}

\subsection{Full spectral fits and kinematic analysis of CO}\label{subsec:fullspec}
We first examine the channel map of the ALMA datacube in Figure \ref{fig:coChanMAP}. The velocities of the channels are measured relative to the rest-frame frequency, assuming $z=2.169$ from \citep{Ishikawa2024}. We see both the extended and compact CO ring structures inferred from the integrated map in Figure \ref{fig:almaCO43}, but the channel maps indicate a connection between the gas morphology and the velocity shifts. The redshifted region extends to the southeast, and the blueshifted region extends into the northwest. Interestingly, it appears that the CO emission associated with the larger velocity offsets $|\Delta v|>150\ \textrm{km s}^{-1}$ is more extended, possibly associated with \targSW, whereas the emission with smaller velocity offsets $|\Delta v|<50\ \textrm{km s}^{-1}$ is more compact, forming the CO ring that connects both quasars.  

Perhaps the most intriguing discovery from the integrated map in Figure \ref{fig:almaCO43} is the compact CO ring within about $\pm 50\ \textrm{km s}^{-1}$ of the systemic redshift. The channel maps suggest a possible velocity differential across the CO ring. For example, the southeast region is brightest in the redshifted gas (up to $100\ \textrm{km s}^{-1}$) with little blueshifted gas, whereas the northwest region is brighter in the blueshifted gas (up to $-150\ \textrm{km s}^{-1}$) with little redshifted gas. 

In Figure \ref{fig:aperspect} we show the aperture spectra extracted from select regions. They are (a) two $r=0.1''$ apertures co-spatial of the two quasars and the CO ring - \targSW\ and \targNE; (b) two additional $r=0.1''$ apertures along the CO ring -  R1 and R2; (c) one $r=0.05''$ aperture in the middle of the CO ring - Center; and (d) three $r=0.15''$ apertures that capture the extended CO emission - E1, E2, and E3. Each spectrum is freely fit with up to three Gaussian profiles. First, we compare the aperture spectra taken at \targSW\ and \targNE. We find that the CO spectra of \targSW\ and \targNE\ are distinctly different. \targSW\ exhibits broad redshifted and blueshifted components, while \targNE\ is dominated by a broad blueshifted component. The broad components have $\sigma\sim 120\pm10\ \textrm{km s}^{-1}$. Due to the large aperture extraction, it is unclear whether the broadening is due to fast-moving gas or is a blend of multiple narrow line components. In this step, we assume a single broad line. Interestingly, \fQratio\ of the integrated CO spectrum is $\sim1.2$, which differs from the continuum ratio of $\sim2.5$. We also examine the CO ring with three additional apertures, R1, R2, and Center. There appears to be a narrow component shared in the inner region (\targSW, \targNE, Center, and R1) with a small, yet varying, velocity offset. The narrow component has widths $\sigma\sim 50\pm10\ \textrm{km s}^{-1}$. R2 does not have an obvious narrow component; however, it is possible that the narrow component and the redshifted, broad components are blended like in the \targNE\ aperture. Center shows a velocity gradient with a narrow redshifted and a broadened blueshifted component. The extended emission at E1, E2, and E3 are narrower at $\sigma\sim 70\pm5\ \textrm{km s}^{-1}$ with some possible blending of the broad components. 

To better understand the kinematics of the CO gas, we fit the spectra at each spaxel across the entire field and decompose CO emissions by their different velocity components. For each spaxel element, we extract and fit the CO spectrum with one or two Gaussians with variable $\sigma$. An additional third Gaussian component was considered, but was not necessary. We discuss the fit results for the single-component and two-component line fits. 

First, we assume the CO emission only requires a single Gaussian component at each spatial element. The corresponding moment maps (integrated line flux $F_{int}$, velocity shift $v_{50}$, and line width $w_{80}$) are shown in Figure \ref{fig:n1fits}. We find that the $v_{50}$ map has a similar kinematic morphology to that of the rest-frame optical (i.e.~ionized \ha) emission lines seen with JWST \citep{Ishikawa2024}. The redshifted region extends to the southeast, and the blueshifted region extends into the northwest. However, unlike the ionized gas velocity map \citep{Ishikawa2024}, the CO velocity map appears more asymmetric and kinematically disturbed. For instance, \targNE\ is surrounded by blueshifted CO, whereas \targSW\ is surrounded by both redshifted and blueshifted gas. Instead of showing the $\sigma$ maps, we show $w_{80}$ that best identifies regions with complex line profiles, likely requiring additional Gaussian components. The $w_{80}$ map in Figure \ref{fig:n1fits} reveals two zones: one with large $w_{80}$ values that reach 340 to 480 \kms\ in dark orange-brown and small $w_{80}$ values with $w_{80}<150\ \textrm{km s}^{-1}$ in light yellow-orange. Regions with large $w_{80}$ coincidentally trace the CO ring noted earlier.%, which correspond to apertures R1, R2, \targSW, and \targNE\ in Figure \ref{fig:aperspect}. %The outer regions with smaller $w_{80}$ values correspond to apertures E1, E2, and E3. 

The measurement of non-uniform distribution and large $w_{80}$ values suggest the need for multiple components, as was demonstrated in Section \ref{subsec:spectanaly}. Next, we examine the two-component CO line fit results. In Figure \ref{fig:n2fits}, we show the moment maps ($F_{int}$, $v_{50}$, and $w_{80}$) for the two-component ($c_1$, $c_2$) fits. We sort the line components by the velocity dispersion, such that $c_1$ traces the narrow component ($\sigma<100\ \textrm{km s}^{-1}$) with a mean $w_{80}$ of $150\pm40\ \textrm{km s}^{-1}$ and $c_2$ traces the broader line component ($\sigma>100\ \textrm{km s}^{-1}$) with a mean $w_{80}$ of $320\pm90\ \textrm{km s}^{-1}$. Line fits with poor $\chi^2$ goodness-of-fit were excluded. 

A different picture of \target\ with kinematically complex CO emerges from the two-component fits. First, distribution of the two gas components is distinct. The $c_1$ component is spatially extended , whereas the $c_2$ component is more spatially compact. Second, the two components have different kinematic properties. The $c_1$ component has complex kinematics with a prominent velocity gradient from $+200\ \textrm{km s}^{-1}$ to $-350\ \textrm{km s}^{-1}$ around \targSW, while the $c_1$ component gas around \targNE\ is mostly blueshifted ranging from $-20$ to $-150\ \textrm{km s}^{-1}$. The observed $v_{50}$ velocity gradients of $c_1$ have a similar asymmetric morphology to that of the single-component fit in Figure \ref{fig:n1fits}. The $c_2$ component is almost entirely blueshifted on average by $-120\pm90\ \textrm{km s}^{-1}$. Third, and most importantly, the kinematic decomposition of the CO emission eliminates the CO ring initially identified by the integrated maps in Figures \ref{fig:almaCO43} and \ref{fig:n1fits}. There is no obvious ring pattern in the integrated flux maps of either component, which may suggest that the CO ring seen in Figure \ref{fig:almaCO43} and \ref{fig:n1fits} is not a real physical feature, but rather a visual artifact due to the blending of multiple kinematically independent emission line components. From the two-component fits of the CO emission, it is clear that the CO gas in \target\ is kinematically disturbed.

\section{Discussion}\label{sec:discuss}
Despite the comprehensive and exhaustive multi-wavelength analysis of \target, differentiating between a physical dual quasar and a pair of lensed images in this target has been challenging. Although this comparison was discussed in detail in \cite{ChenYC2023a, Ishikawa2024}, we address the lensing hypothesis in the context of the new ALMA results. Then we discuss the implications of the observed \coline\ and continuum observations. 

\subsection{Making the dual quasar case}\label{sec:DISC:lens}
We summarize all key arguments for/against the lensing and dual quasar hypotheses in Table \ref{tab:lensVdual} from this study and past observations \citep{Shen2021, ChenYC2022, ChenYC2023a, ChenYC2024, Ishikawa2024}. 
There are some predictions from the two scenarios that we can compare the data with. In a lensed system, the pair of lensed images of the quasar and its host galaxy would lie along a thin Einstein Ring. Although some spectral variations may be expected, the quasar images would have near-identical spectra with little chromatic dependence. The ``smoking-gun'' evidence against the lens would be the detection of distinctly different spectra - not explainable by differential reddening - and a velocity offset between the two quasars. 

\begin{table*}[t]
\caption{We summarize all known evidence (morphological, photometric, spectroscopic, and kinematic) for/against the lensing and dual quasar hypotheses. Despite some strong evidence for the lensing hypothesis, they are contradicted by evidence in favor of the dual quasar hypothesis. With the latest addition of ALMA data, we argue that \target\ is indeed a dual quasar. } 
\label{tab:lensVdual} 
\begin{tabular}{lccc}
    \hline
    Evidence & Data  & Lensed Quasar & Dual Quasar \\
    \hline
    %%%%%%%%%%%%%%%%%%%%%%%%%%%%%%%%%%%%%%%%%%%%%%%%%%%%%%%%%%%%%%%%%%%%%%%%
    Discovery of a quasar pair & {a, b} & 
    $\phantom{^*}\triangle\phantom{^*}$ & $\phantom{^*}\triangle\phantom{^*}$ \\
    
    No detection of foreground lens galaxy & {b, d} &  
    $\phantom{^*}\triangle\phantom{^*}$ & $\phantom{^*}\triangle\phantom{^*}$ \\
    
    Detection of tidal tails & {b, d}  &  
    $\phantom{^*}-\phantom{^*}$ &  $\phantom{^*}\bigcirc\phantom{^*}$ \\
    Detection of extended multi-phase emission (e.g.~\ha, PAH, CO) & {c, d, e}  &  $\phantom{^*}-\phantom{^*}$ &  $\phantom{^*}\bigcirc\phantom{^*}$ \\
    
    %Detection of extended \coline\ gas & {E} &  $\phantom{^*}-\phantom{^*}$ &  $\phantom{^*}\bigcirc\phantom{^*}$ \\
    %\fQratio\ consistently $\sim3$ at select bands & {B, D, E} & 
    %$\phantom{^*}\triangle^*$ & $\phantom{^*}-\phantom{^*}$ \\
    
    $\lambda$-dependent \fQratio\ differences & {b, d, e} &  
    $\phantom{^*}\triangle\phantom{^*}$ & $\phantom{^*}\triangle\phantom{^*}$ \\
    
    Spectral similarity of nuclear optical & {d}  & 
    $\phantom{^*}\bigcirc^*$ & $\phantom{^*}\triangle\phantom{^*}$ \\
    
    Spectral differences of nuclear \coline  & {e} &  
    $\phantom{^*}-\phantom{^*}$ &  $\phantom{^*}\bigcirc^*$ \\
    
    Velocity offset ($\sim200\ \textrm{km s}^{-1}$, optical) between the quasars & {d} & 
    $\phantom{^*}-\phantom{^*}$ & $\phantom{^*}\triangle\phantom{^*}$ \\
    
    Two SMBH with similar properties (\Mbh\ and \Lbol) & {d} & 
    $\phantom{^*}\triangle\phantom{^*}$ & $\phantom{^*}\triangle\phantom{^*}$ \\
    
    \ha\ kinematics suggest rotating gas & {d} &  
    $\phantom{^*}-\phantom{^*}$ &  $\phantom{^*}\bigcirc\phantom{^*}$ \\
    
    Morphologically and kinematically complex CO  & {e} &  
    $\phantom{^*}-\phantom{^*}$ &  $\phantom{^*}\bigcirc\phantom{^*}$ \\
    
    Detection of CO ring & {e} & $\phantom{^*}\triangle^*$  & 
    $\phantom{^*}\triangle\phantom{^*}$ \\
    
    CO ring eliminated with kinematic decomposition & {e} &  
    $\phantom{^*}-\phantom{^*}$ &  $\phantom{^*}\bigcirc^*$ \\
    %%%%%%%%%%%%%%%%%%%%%%%%%%%%%%%%%%%%%%%%%%%%%%%%%%%%%%%%%%%%%%%%%%%%%%%%
    \hline
\end{tabular}
\tablecomments{ $\bigcirc$ Strong evidence that supports the selected hypothesis. $\triangle$ Ambiguous evidence that may support either hypothesis or the measurement is uncertain. $*$ Contradictory evidence. \textsuperscript{a} \cite{ChenYC2022}; \textsuperscript{b} \cite{ChenYC2023a}; \textsuperscript{c} \cite{ChenYC2024}; \textsuperscript{d} \cite{Ishikawa2024}; \textsuperscript{e} This ALMA study.}
\end{table*}   
The most surprising result from JWST/NIRSpec aperture spectroscopy was the spectral similarity of the two quasars \citep{Ishikawa2024}. Furthermore, our ALMA observations reveal a compact CO ring that connects the two quasars as shown in Figures \ref{fig:almaCO43} and \ref{fig:coChanMAP}. This CO ring morphology is reminiscent of a gravitationally lensed Einstein ring at the angular separation of the two quasars. These two observations are the two leading evidence for lensing. 

The key evidence for the dual quasar hypothesis in this study lies in the gas kinematics analysis. First, extended multi-phase gas, reaching beyond the supposed Einstein radius, has been detected with both JWST \citep{Ishikawa2024} and ALMA (Figures \ref{fig:cont}-\ref{fig:coChanMAP}). This contradicts the expectations of a lensed Einstein ring, which typically has narrow ring widths, in which most of the CO emission should be confined to within the ring. Second, the multi-phase emission corresponding to the two quasars is not identical; we do not see the two mirror images expected from a lensed quasar. JWST detected subtle differences in the optical emission line profiles, including a velocity offset of $\sim200\ \textrm{km s}^{-1}$ between the two quasars \citep{Ishikawa2024}. Aperture spectra taken along the CO ring, \targSW, and \targNE\ in Figure \ref{fig:aperspect} show significant differences in the spectral profiles, including significant variations in the underlying spectral components. We require different spectral fits for each region: \targSW\ requires three components (two red-/blue-shifted broad components and one narrow component), whereas \targNE\ only requires two blueshifted components (one broad component and one narrow component). Furthermore, ALMA shows that the spatial distribution of the brightest CO clumps does not align with the known quasar positions (contours in Figure \ref{fig:almaCO43}), which is inconsistent with lensing geometry. While spectral variations in the extended features are not surprising, spectral differences of spectra taken from apertures co-spatial with the two quasars are more suggestive of different physical properties than lensed images. 

Third, spatially resolved spectroscopy revealed a complex dynamical environment. The kinematic $v_{50}$ maps of both the ionized \ha\ \citep{Ishikawa2024} and molecular gas (this study) revealed velocity differentials that are suggestive of a large rotating disk or an ongoing merger.  Furthermore, the expected parity of the red-/blue-shifted regions is not consistent with a lens. If the strongly lensed quasar host galaxy is a disk, then we expect to see a specific kinematic pattern (a mirror pair of red/blue-shifted velocity gradient along the Einstein ring, such as in \citealt{Riechers2008} and \citealt{LiuD2024}). Instead, the $v_{50}$ maps in Figures \ref{fig:n1fits} and \ref{fig:n2fits} contradicts this lensing prediction. Most importantly, the kinematic decomposition of the CO emission in Section \ref{subsec:fullspec} eliminates the CO ring as shown in Figure \ref{fig:n2fits}. If the CO ring is an Einstein ring, the kinematic decomposed intensity maps and $v_{50}$ maps should still trace the Einstein ring, which is not what we see. Rather, the CO ring may either represent a real physical phenomenon, such as nuclear rings \citep[e.g.][]{Stuber2023}, or a visual artifact due to the superposition of multiple kinematically independent emission line components.

Other evidence is ambiguous and less robust. In addition to the previously measured wavelength-dependent flux differences \citep{ChenYC2023a, Ishikawa2024}, \fQratio\ measured with ALMA - the 455 GHz continuum ($\sim2.5$) and the 650 GHz CO emission ($\sim1$) - differ. Flux variations likely arise from real physical processes, but they may also result from lensing effects, such as differential reddening \citep[e.g.,][]{Sluse2012}, perturbations of the lens mass model \citep[e.g.,][]{Stacey2018}, or misalignment between the lens and source \citep{Agnello2018}. Complex gas kinematics have been seen in some lensed quasars \citep[e.g.][]{Riechers2008, Stacey2018, Lamperti2021}. Also, the non-detection of a foreground lens \citep{ChenYC2022, ChenYC2024, Ishikawa2024} does not necessarily confirm the absence of a lens (see \citealt{Hawkins1997a}). 

If \target\ is a lensed quasar, then the required lens configuration is two quasar images, surrounded by extended ionized and molecular gases. One possibility is that only the circumnuclear region is lensed and the extended gas appears unlensed because the nuclear gas is depleted. However, this may be an unlikely scenario; either the quasar host galaxy needs to be extremely large in extent or the undetected foreground lens galaxy is extremely massive and compact.

\begin{figure*}
    \begin{center} 
     \includegraphics[width=0.67\textwidth,trim={0.4cm 0 0.3cm 0},clip]{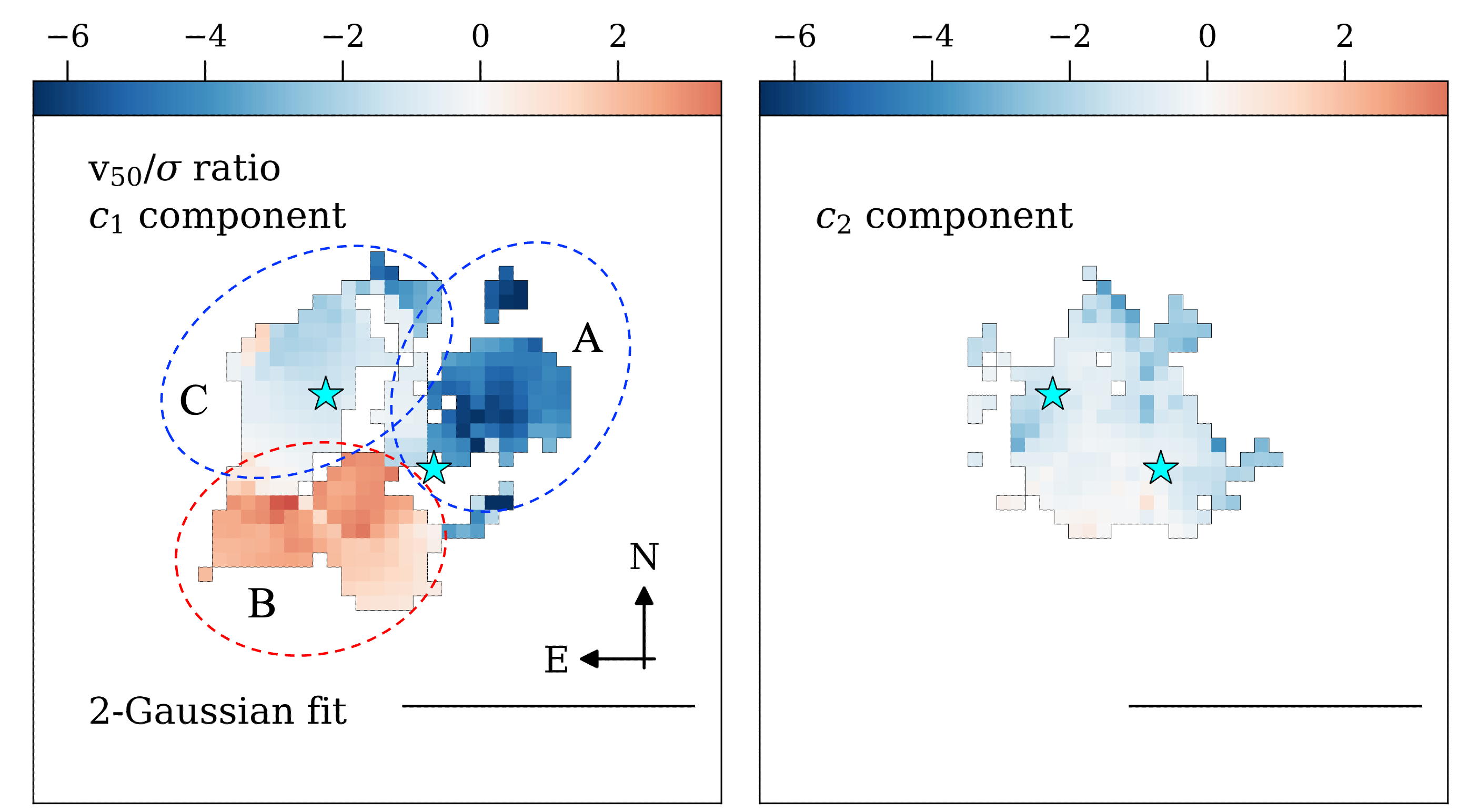}
     \end{center}
    \caption{We show the $v_{50}/\sigma$ ratio of the two-component fits from Figure \ref{fig:n2fits}. The red and blue colors indicate redshifted and blueshifted gas, respectively. (Left) The $c_1$ component, tracing $\sigma\sim50\ \textrm{km s}^{-1}$, is kinematically complex gas. It has three distinct kinematic regions: (A) a strongly blueshifted gas with large $v_{50}/\sigma\sim-5$, (B) a moderately redshifted gas with $v_{50}/\sigma\sim2.5$, and (C) a moderately blueshifted gas with $v_{50}/\sigma\sim-1$. (Right) The $c_2$ component, tracing $\sigma\sim150\ \textrm{km s}^{-1}$, appears to be dispersion dominated, likely due to turbulent fast-moving gas, possibly triggered by the merger dynamics.}
    \label{fig:n2vsig} 
\end{figure*}

As summarized in Table \ref{tab:lensVdual}, we learn that ruling out the lensing scenario is not straightforward. However, based on the kinematic analyses, we make the case that \target\ is a dual quasar rather than a lensed double quasar. The challenge with the existing JWST and ALMA data is that we only detect emission lines from the gas associated with the quasar host galaxy. We have yet to detect the stellar continuum, which may better reveal merger signatures. The difficulty may also be characteristic of small-separation dual quasar candidates like \target. In the following sections, we discuss the implications of the observed molecular gas properties, assuming the dual quasar interpretation. %We will explore the interpretation of the CO dynamics in detail in Section \ref{sec:DISC:co}.

\newpage
\subsection{Interpreting the CO kinematics}\label{sec:DISC:co}
The emerging picture of \target\ is a dual quasar system with similar SMBH properties embedded in a kinematically complex ionized and molecular gas system. In Section \ref{subsec:fullspec}, we presented the spectral fits of a kinematically disturbed CO. The emission line fits shown in Figures \ref{fig:aperspect} and \ref{fig:n2fits} suggest that \target\ is made up of at least two kinematically different gas components: an extended gas component traced with narrow lines ($w_{80}\sim150\ \textrm{km s}^{-1}$) and a compact gas component traced with a broad lines  ($w_{80}\sim320\ \textrm{km s}^{-1}$). %This picture is consistent with a major merger.  

Based on the line widths of both $c_1$ and $c_2$ CO components, it is unlikely that the blue-/red-shifted $v_{50}$ velocity gradient traces biconical quasar-driven molecular outflows. Some quasar-driven molecular outflows are typically characterized by extremely broad lines with FWHM reaching $500-1,000\ \textrm{km s}^{-1}$ \citep[e.g.][]{Stacey2022}. Although the regions around the two quasars show broader line widths reaching $w_{80}\sim500\ \textrm{km s}^{-1}$, the majority of the broad line gas is dominated by $w_{80}\lesssim400\textrm{ km s}^{-1}$. Thus, neither component of the two Gaussian fits meets the quasar-outflow criterion. However, we cannot rule out the presence of slow-moving molecular outflows. Interestingly, no ionized gas outflows have been observed with JWST either \citep{Ishikawa2024}, indicating that quasar feedback may not yet be dominant in \target. If the observed gas kinematics is not well explained by fast-moving quasar outflows, then alternative explanations are merger-driven dynamics or circumnuclear gas flows. 

A notable feature of the narrow $c_1$ component is the velocity gradient in its $v_{50}$ map (Figure \ref{fig:n2fits}). This velocity gradient may be evidence for rotation; however, the asymmetry of the $v_{50}$ map complicates this interpretation \citep{Wisnioski2015}. This asymmetry contradicts the analysis of the ionized gas kinematics seen with JWST, which suggested the presence of a large rotating gas disk enveloping both quasars \citep{Ishikawa2024}. Alternatively, the asymmetry of the CO gas kinematics may represent non-circular motions. 

In Figure \ref{fig:n2vsig} we map the \vsigRat\ ratio corresponding to each of the CO components to characterize the dynamical state of the molecular gas. The \vsigRat\ maps suggest a kinematically disturbed environment with two dynamically distinct systems: a dynamically complex, slow-moving gas and a turbulent, fast-moving gas. 

The slow-moving $c_1$ component appears to be a mix of different dynamical states: a strongly blueshifted gas with large $v_{50}/\sigma\sim-5$ (region A), a moderately redshifted gas with $v_{50}/\sigma\sim2.5$ (region B), and a moderately blueshifted gas with $v_{50}/\sigma\sim-1$ (region C). The non-uniform distribution of \vsigRat\ indicates that the $c_1$ component is likely not dominated by large-scale rotation. However, since we detect the positive and negative gradient in $v_{50}$ with relatively large \vsigRat\ centered on \targSW\ (regions A and B), we cannot rule out the presence of a counter-rotating gas, such as a disky host galaxy of \targSW. Due to the sharp change in $v_{50}/\sigma$ along the boundary of regions B and C, these two regions are not likely to represent rotation. Thus, we have two plausible interpretations of the $c_1$ component: a gas rotation that is kinematically disturbed by the merger, kinematically misaligned and turbulent gas structures, or a combination of both. Numerical simulations also predict that the host galaxies of dual quasars may be disk-dominated \citep{Dadiani2024}. The presence of rotation-like kinematics may indicate that \target\ is still in the early stages of the galaxy merger. On the other hand, the $c_2$ component has a more uniform distribution of $v_{50}/\sigma\sim-1$, which suggests a dispersion-dominated dynamics. The larger velocity dispersion of the $c_2$ component also suggests turbulence, possibly heated by a merger, in which this CO gas phase has yet to settle into ordered motion. 

The detection of two kinematically complex and distinct CO components suggests the presence of a dual-phase dynamical system that is spatially overlapped. The broadened and blueshifted $c_2$ gas component suggests the presence of turbulent and fast-moving gas, moving toward the observer. However, it is unclear if the $c_1$ component traces a rotating gas enveloping both quasars or if it traces two separate galaxies in a merger (i.e. a rotating host galaxy of \targSW\ and a turbulent host galaxy of \targNE\ that is moving towards the observer). Furthermore, it is unclear from the kinematic maps how or if the $c_2$ gas component is coupled to the two quasars. In Figure \ref{fig:n2fits}, there is a small velocity gradient of $\Delta v\sim120\ \textrm{km s}^{-1}$ from \targNE\ to \targSW, so $c_2$ gas may trace a tidal bridge connecting \targNE\ and \targSW, similar to one observed by \cite{Izumi2024}. One hypothesis also presented in \cite{Ishikawa2024} is two merging host galaxies that appear superimposed to the observer \citep[e.g.][]{Tubin2021, Ciraulo2021}, which results in kinematic ambiguity at the current spectral and spatial resolutions. 

Despite the ambiguity in the exact dynamical state of \target, based on the kinematic analysis of the CO gas it is clear that \target\ is likely in an ongoing galaxy merger that fuels the intense star formation \citep{ChenYC2024, Ishikawa2024} and dual quasar activity. Two possible ways to improve the dynamical assessment in the future is to compare the kinematics of the stellar continuum with the ionized and molecular gas components \citep{Wisnioski2015} and also to obtain higher spatial resolutions with ALMA. 

\subsection{Molecular mass estimates}\label{sec:DISC:Mmol}
\begin{table*}
\centering
\caption{We show the total molecular mass, \Mmol, derived from $L'_{CO}$ listed in Table \ref{tab:CO43}. We calculate \Mmol\ using different CO excitation correction from \coline\ to \cobase, $R_{41}$, and different CO-\hmol\ conversion factors, $\alpha_{CO}$. Our baseline calculation assumes an optically thick, thermalized gas $R_{4,1}\approx1$ with $\alpha_{CO}=0.8$, corresponding to local dusty starbursts and quasars. $R_{4,1}=0.87$ is typically used for local quasars, and $R_{4,1}=0.37$ corresponds to the calibration of IR-selected AGNs at cosmic noon \citep{Kirkpatrick2019}. We place an upper limit on \Mmol\ with $\alpha_{CO}=3.2$, corresponding to main-sequence galaxies. The mean mass for the different relations is $\langle M_{H_2} \rangle = (6\pm 5)\times10^{10} M_{\odot}$.} 
\label{tab:Mmol} 
    \begin{tabular}{cccc}
    \hline
      & $R_{4,1}=0.37 \pm 0.11$ &  $R_{4,1}=0.87$ &  $R_{4,1}=1$ \\
    \hline
    $\alpha_{CO}=0.8$ 
        & $(3.9 \pm 0.1) \times 10^{10}M_{\odot}$ & $(1.8 \pm 0.1) \times 10^{10}M_{\odot}$   & $(1.6 \pm 0.1) \times 10^{10}M_{\odot}$ \\
    $\alpha_{CO}=3.2$ 
        & $(1.6 \pm 0.1) \times 10^{11}M_{\odot}$ & $(7.1 \pm 0.5) \times 10^{10}M_{\odot}$   & $(6.2 \pm 0.4) \times 10^{10}M_{\odot}$  \\
    \hline
    \end{tabular}
\end{table*}
Since CO traces \hmol\ \citep{Carilli2013}, we estimate the bulk \hmol\ mass, \Mmol, using the $M_{H_2}=\alpha_{CO}L'_{CO(1-0)}$ relation. Since the CO-\Mmol\ conversion is based on $L'_{CO(1-0)}$ for the \cobase\ transition, we convert from $L'_{CO(4-3)}$ (Table \ref{tab:CO43}) with the following assumptions. First, we assume that the low-J CO transitions are thermalized and optically thick using the CO excitation correction of $R_{41}=L'_{CO(4-3)}/ L'_{CO(1-0)} \approx 1$, as in \cite{Vayner2021a}. Following the \cite{Bertola2024} analysis, we also consider $R_{\textrm{41}}=0.87$, determined from local quasars \citep{Carilli2013}, and $R_{\textrm{41}}=0.37\pm0.11$, determined from IR-selected AGNs at cosmic noon \citep{Kirkpatrick2019}. Second, we adopt the CO-\hmol\ conversion factor of $\alpha_{CO}=0.8\ \textrm{K km s}^{-1}\textrm{ pc}^2$, typical for local quasars and ULIRGs \citep{Solomon2005, Tacconi2008}, although there are many uncertainties of at least 30\% \citep{Bolatto2013, Papadopoulos2012} that depend on different conditions in the interstellar medium like gas metallicity to host galaxy's relation to the main-sequence \citep{Elbaz2007, Noeske2007, Accurso2017}. We also consider $\alpha_{CO}=3.2$ for main-sequence galaxies to place an upper limit on \Mmol. Under these assumptions, we calculate \Mmol\ with the following relation:
\begin{equation}
    M_{H_2} = \alpha_{CO}R_{41}^{-1}L'_{CO(4-3)}.
\label{eq:MH2CO}
\end{equation}

We summarize the \Mmol\ calculations in Table \ref{tab:Mmol}. Our baseline \Mmol\ estimate for $R_{\textrm{41}}=1$ is $M_{H_2}=(1.6\pm0.1)\times10^{10}\  \textrm{M}_{\odot}$. Depending on our choice of $R_{\textrm{41}}$ and $\alpha_{CO}$, we obtain different estimates for \Mmol. Although the choice of $R_{\textrm{41}}$ only results in variations of a factor of few, the choice of $\alpha_{CO}$ results in a variation of about 0.5 dex. The upper limit to \Mmol\ is $M_{H_2}=(1.6\pm0.1)\times10^{11}\ \textrm{M}_{\odot}$, based on the $\alpha_{CO}=3.2$. The mean value over all of the estimates is $\langle M_{H_2}\rangle=(6\pm5)\times10^{10}\ \textrm{M}_{\odot}$. 
 
Based on our initial \Mmol\ calculations, \cite{ChenYC2024} compared $M_{H_2}$ with the \pah-based SFR of $\sim1,000\ M_{\odot}\textrm{ yr}^{-1}$ and found that \target\ is an outlier with an extreme starburst rate that exceeds the available fuel, according to the Kennicutt-Schmidt relation (\citealt{Kennicutt1998}). The relative agreement with the Kennicutt-Schmidt relation depends strongly on the choice of $R_{\textrm{41}}$ and $\alpha_{CO}$. For example, if we choose values corresponding to local quasars and dusty starbursts, \target\ deviates strongly from the Kennicutt-Schmidt relation. However, if \target's host galaxy is a main-sequence galaxy, then the calculated SFR/\Mmol\ ratio is consistent with the Kennicutt-Schmidt relation. Based on the known star formation and quasar properties of \target\ and the inferred merger activity, it is unlikely that its host galaxy is main-sequence-like.

One explanation of the deviation from the Kennicutt-Schmidt relation is that the merger dynamics help facilitate the rapid conversion of gas to stars, as well as fueling of the two quasars. An alternative interpretation is that the host galaxy environment of \target\ has an extended starburst phase despite the depletion of the molecular gas reservoir \citep{ChenYC2024}. The high star formation efficiency is similar to hyperlumious quasars at cosmic noon \citep{Bischetti2018}. Although quasar-feedback is known to cause gas depletion \citep[e.g.][]{Bertola2024}, fast gas outflows have not been detected in the molecular (this study) or ionized phases (JWST; \citealt{Ishikawa2024}). Thus, quasar feedback may be less impactful in \target\ unlike some dual quasars seen by \cite{Tang2024}. \target\ provides evidence for enhanced star formation, concurrent with quasar activity, in a major gas-rich galaxy merger.

\begin{figure*}
 \begin{center}
    \begin{tabular}{cc}
    \includegraphics[width=0.5\textwidth]{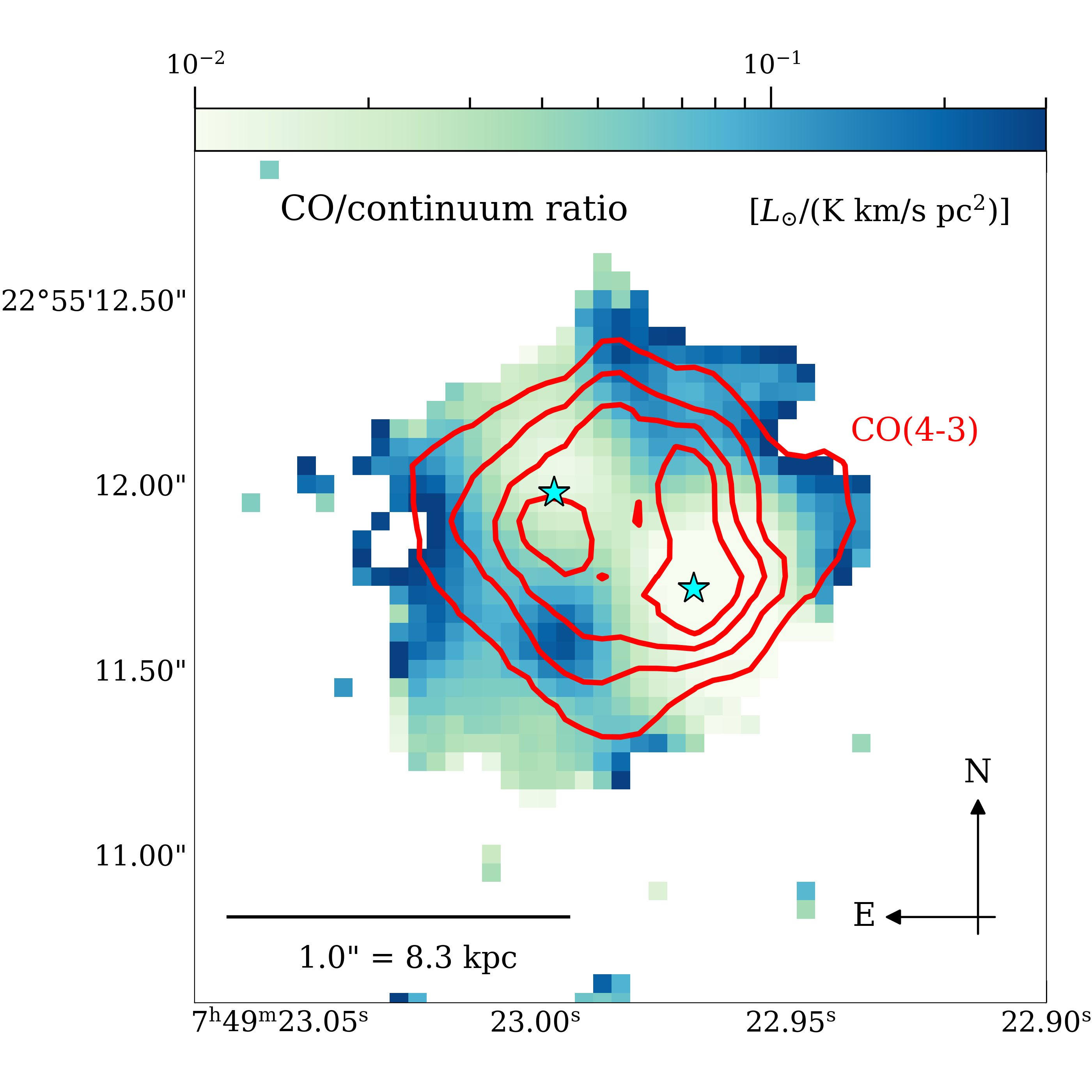} &
    \includegraphics[width=0.45\textwidth]{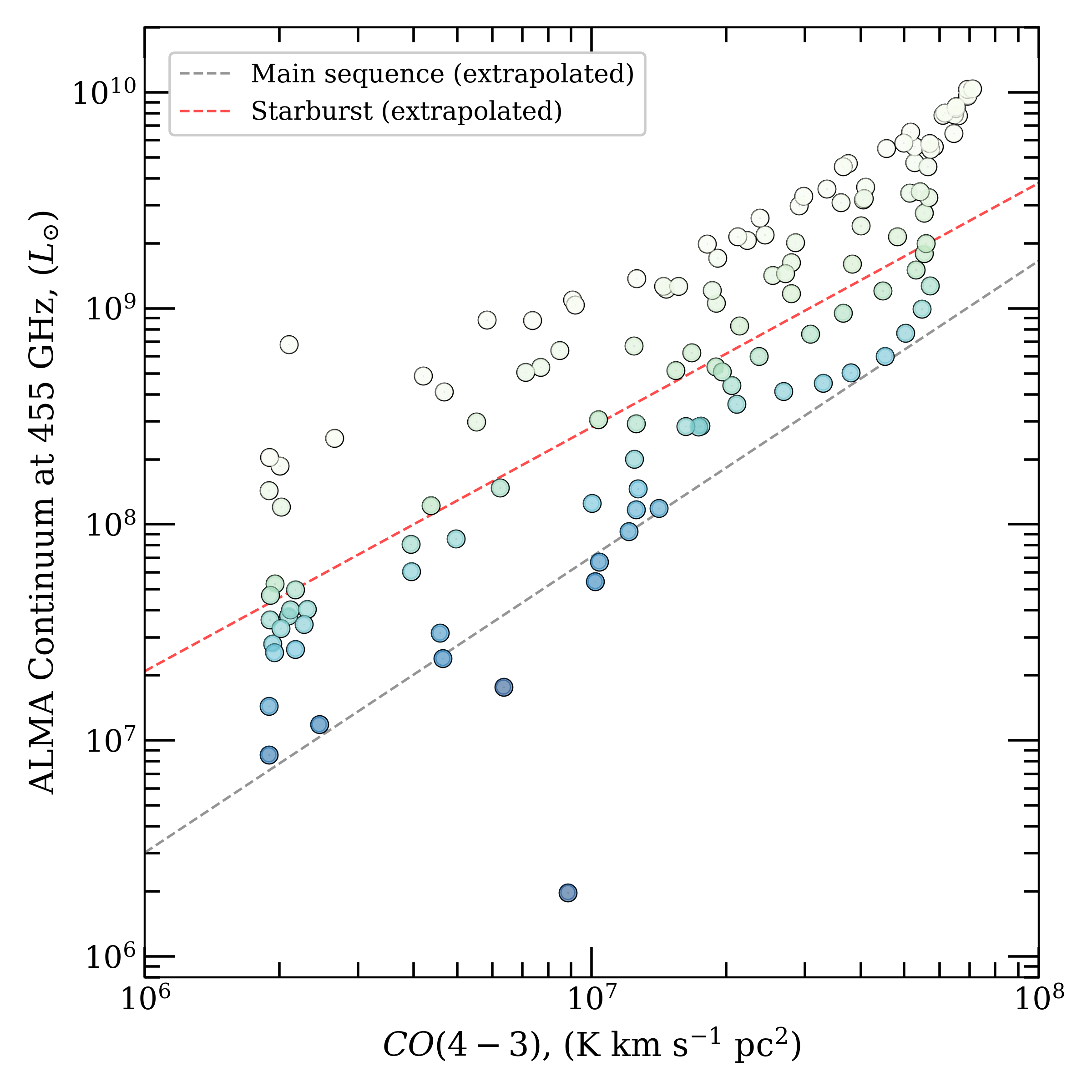} 
    \end{tabular}
     \end{center}
	 \caption{(Left) Map of the CO/continuum emission ratio. The red contours outline the clumpy structure of the CO emission at $\textrm{SNR}>2$. We can see that the light-green regions, dominated by the nuclear synchrotron emission, have an elevated continuum despite the presence of \coline, traced with the red contours. The outer regions of elevated CO, away from the continuum, in dark blue are likely dominated by star formation. (Right) The CO/continuum ratio separated by the  differential spatial elements. The color shading matches the CO/continuum map. We compare to the known relations of local starburst and main-sequence galaxies (\citealt{Carilli2013} and references therein).} 
	 \label{fig:contCO} 
\end{figure*}

\begin{figure*}
	 \begin{center}
    \includegraphics[width=0.8\textwidth,trim={0.2cm 0.4cm 0.2cm 0.2cm},clip]{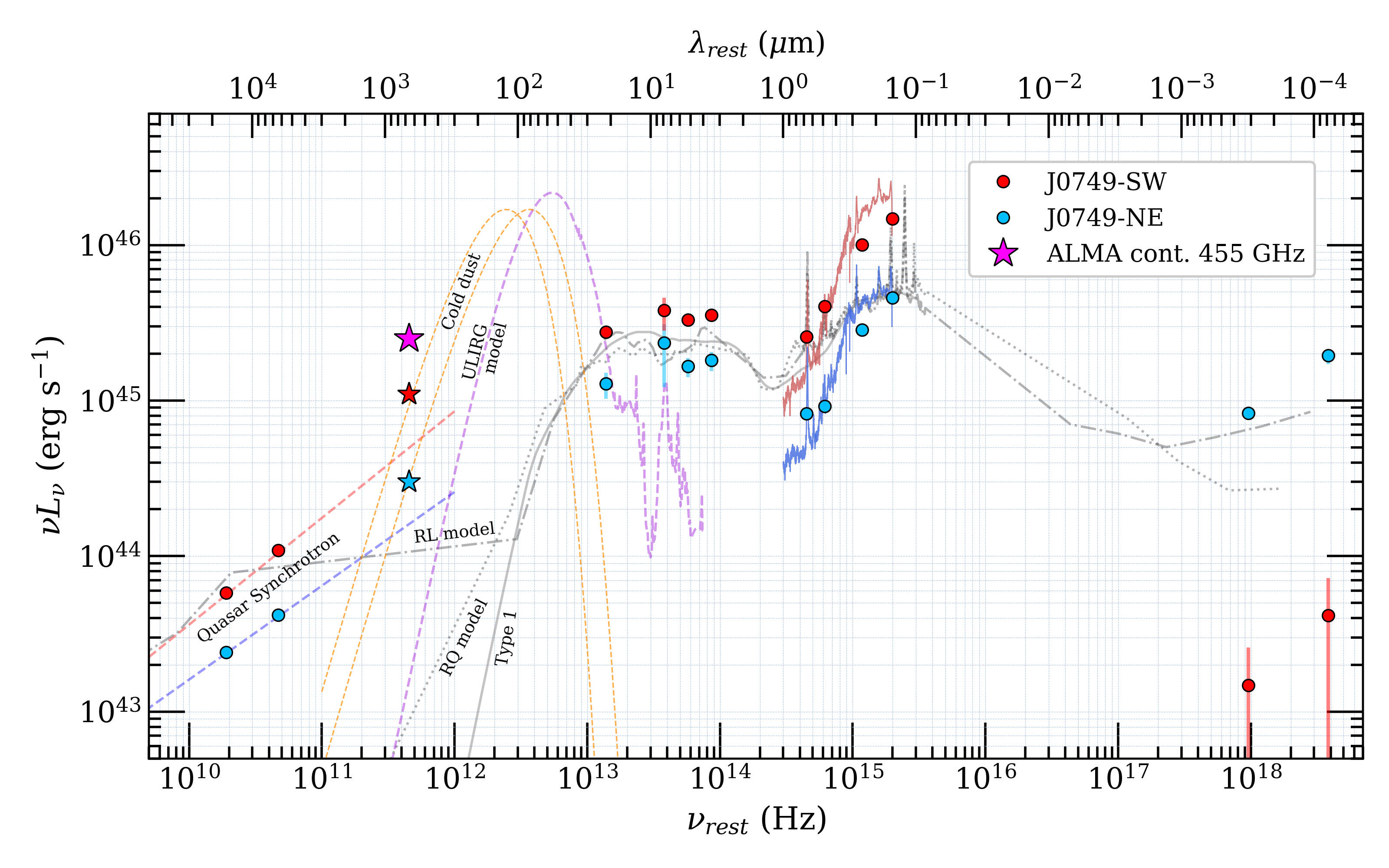}
	 \end{center}
	 \caption{An updated SED of \target\ adapted from \cite{ChenYC2023a,ChenYC2024} with the new ALMA measurements of the rest-frame 455 GHz continuum, optical and near-infrared spectroscopy \citep{Ishikawa2024}. The circles distinguish each quasar by color: \targSW\ (red), \targNE\ (blue), and integrated (purple). The radio-loud and radio-quiet SED models (grey dotted and dashed lines) are from the \cite{Shang2011} templates. The light purple dashed line shows the ULIRG template \citep{Rieke2009, Bell2003} calibrated to the known SFR. The radio/sub-mm luminosity converted from SFR is too small to be shown in this plot. The light red and blue dashed lines show the extrapolated synchrotron continuum. We also show two blackbody spectra assuming cold dust in dashed light orange.} 
	 \label{fig:sed} 
\end{figure*}

\subsection{Synchrotron dominated continuum}\label{sec:DISC:cont}
The bright 455 GHz continuum is co-spatial with the bright radio sources seen in VLA \citep{ChenYC2023a}. In Figure \ref{fig:contCO}, we map and plot the flux ratio values of the observed \coline\ and continuum, CO/continuum. We see a deficit of CO in the nuclei and an enhancement in the outskirts. This suggests that the nuclear component may dominate the observed continuum, which complicates the original goal of estimating the total far-infrared (FIR) luminosity to calculate SFR. 

We also compare CO/continuum against the known $L_{\textrm{TIR}}$ vs.~CO ratio \citep{Carilli2013} in Figure \ref{fig:contCO}. The caveat is that this is not a direct comparison since \cite{Carilli2013} compares a galaxy's global properties, whereas we can do this on a spaxel-by-spaxel basis within one object. Assuming that this relation is true at small spatial scales, $<1\ \textrm{kpc}$, we extrapolate the \cite{Carilli2013} relation to lower $L_{\textrm{TIR}}$ and $L_{\textrm{CO}}$ values. It appears that the central regions have CO/continuum values that are at least comparable to or greater than starbursts, supporting the interpretation of enhanced star formation in \target. 

We also examine the continuum emission in the context of the two quasars' spectral energy distribution (SED). From the \ha\ and \pah\ emission \citep{ChenYC2024, Ishikawa2024}, we know that \target\ is hosted by an extreme starburst ($\textrm{SFR}>1,000\ M_{\odot}\textrm{ yr}^{-1}$). Since a quasar hosted by an extreme starburst is analogous to ultra-luminous infrared galaxies (ULIRGs), we compare the SED of \target\ with the template SED of ULIRGs \citep{Rieke2009}, scaled by the known SFR of \target. SFR is known to be correlated with the total infrared luminosity, $L_{\textrm{TIR}}$, summed over 8-1000 \mum. We estimate to be $L_{\textrm{TIR}}\approx6\times10^{12}\ L_{\odot}$ using the \cite{Bell2003} calibration.
In Figure \ref{fig:sed} we show the updated SED of \target\ with the integrated continuum luminosity from Table \ref{tab:cont}. The SED also includes broadband photometric and spectroscopic data, including the most recent JWST NIRSpec and MIRI observations \citep{ChenYC2022, ChenYC2023a, ChenYC2024, Ishikawa2024}. We compare the observations with the quasar synchrotron power-law spectrum  ($f_{\nu}\propto \nu^{-\alpha}$) extrapolated to 1 MHz, the model ULIRG starburst spectrum, a model Type 1 quasar FIR continuum \citep{Lyu2017}, and model radio-loud/quiet quasars \citep{Shang2011}. The starburst's contribution to the radio and sub-mm is also negligible ($\sim10^{41}\ \textrm{erg s}^{-1}$ at 455 GHz). We see that neither the starburst nor the synchrotron model sufficiently explains the observed ALMA continuum. It appears that ALMA detects an excess sub-mm emission at 455 GHz. 

The question of excess sub-mm- and mm-excess emission has been studied in radio-quiet quasars \citep[e.g.][]{Panessa2019, Hermelo2016}. These systems typically have a flat or inverted synchrotron slope with an excess radio emission at higher frequencies. From \cite{ChenYC2023a}, it is known that \target\ is associated with two compact radio luminous cores with a loudness parameter of $R>600$ and flat or inverted radio spectra. Likely explanations of the excess are optically thick (unlikely) synchrotron from a compact jet, magnetically heated corona from the inner-accretion disk, thermal free-free emission from the narrow line region or starburst activity or outflows (\citealt{Panessa2022}, and references therein). 

Alternatively, the observed sub-mm continuum may arise from excess blackbody dust emission. We calculate the blackbody spectrum for cold dust with temperatures $T\approx30\ \textrm{K}$ and $T\approx45\ \textrm{K}$, normalized by the SFR-predicted IR luminosity, $L_{\textrm{TIR}}\sim10^{12}\ L_{\odot}$. We plot both spectra in Figure \ref{fig:sed}. This dust model assumes a cold dust environment within a $r=1.2\ \textrm{kpc}$ aperture around each quasar, despite potential heating from the luminous quasar radiation, intense starburst, and dynamical heating from merger activity. Also, there are no constraints on the dust properties; the two quasars are optically bright with no Balmer decrements detected \citep{ChenYC2023a, Ishikawa2024}. The combined contribution from the cold dust emission and the excess quasar synchrotron radiation may explain the observed 455 GHz continuum. However, this remains speculative without a constraint on the FIR and dust continuum. %It appears that  \red{Since outflows have not been detected in \target, the likely interpretation is that the galaxy's central regions are photoionized by the quasar, which is consistent with the optical line ratio diagnostics in \cite{Ishikawa2024}. }

\subsection{Comparison and implication}
The JWST \citep{ChenYC2024, Ishikawa2024} and ALMA (this study) observations of \target\ reveal a gas-rich host galaxy with intense star formation. The inferred Kennicutt-Schmidt relation of SFR vs.~molecular mass showed that the molecular mass is lower than expected for the calculated SFR, suggesting an intense starburst rate exceeding the available fuel. The multiphase gas kinematics are also complex. Although the ionized gas appears to suggest a single large rotating gas \citep{Ishikawa2024}, the CO kinematics reveals a dynamically disturbed environment,  likely due to a merger, with two kinematically distinct structures. % - an extended tail and a messy interior. 

If \target\ is a result of a major galaxy merger, then the observed properties are similar to other known extreme starbursts like luminous infrared galaxies (LIRGs) and their extreme siblings, ultra-LIRGs (ULIRGs) and hyper-LIRG (HyLIRGs; \citealt[e.g.][]{Harrington2018,LiuD2024}) at cosmic noon. The SED fit shown in Figure \ref{fig:sed} suggests that \target\ has elevated mm-continuum not explained by starbursts, possibly due to additional contributions from the quasars. We can also see this trend in the CO/continuum ratio in Figure \ref{fig:contCO}. The CO depletion suggested by the CO/continuum ratio is also consistent with luminous quasars at cosmic noon \citep{Bischetti2018, Bertola2024}, likely by extremely efficient star formation rather than quasar feedback. However, \cite{Tang2024} detected large molecular gas reservoirs in low redshift dual quasars.

Interestingly, studies of these extreme starbursts have also revealed rotating gas disks traced with CO emission \citep[e.g.][]{Downes1998}. Assuming the presence of rotating CO gas, according to the kinematic decomposition in Section \ref{subsec:fullspec}, then one interpretation is that \target\ may be a superposition of two overlapping galaxies during an ongoing merger, either by chance alignment with the observer or in a close-passage. The detection of distinct kinematic gas components associated with each quasar may support this hypothesis. There is some evidence for this, possibly among low redshift dual quasars. \cite{Koss2023} detected distinct 231 GHz continuum cores and a velocity differential in the CO(2-1) emission, which serves as evidence for two kinematically independent nuclei in Mrk 463 at $z = 0.005355$. Mrk 739, a $z=0.02985$ dual quasar with $\sim3.4$ kpc is more kinematically ambiguous and has been interpreted with the overlapping-galaxy model \citep{Koss2011, Tubin2021}. There are indications from simulations that dual quasars at $z\sim2$ are most likely hosted by disk host galaxies \citep{Dadiani2024}. The other interpretation of \target\ is that it is undergoing a violent merger. Kinematic analysis of CO(2-1) by \cite{Tang2024} reveals that their sample of local dual quasars ($z\approx0.5$) are merger-like, yet ambiguities with the rotation interpretation remain. Clearly, these studies have the advantage of proximity that enables detailed studies of dual quasars. 

These comparisons suggest that dual quasars have distinct properties from starbursts, although the exact cause is not yet clear. According to the case study of \target, it would seem that mergers may fuel and trigger intense star formation. However, the connection with the quasar activity remains unclear, nor is there a consensus in the literature. \cite{Tang2024} showed that local dual quasars have no obvious correlation between the molecular properties and quasar separation. Quasar feedback was implied in some dual quasars \citep[e.g.][]{Izumi2024, Tang2024, Ruby2024}; however, \target\ appears to show little evidence for quasar feedback. While \target\ may be an outlier, it appears that merger connection with quasar feedback remains an open question. A more robust sample comparing the host galaxies of dual quasars (e.g., SFR, dynamics, gas content) and galaxy mergers is needed to fully address the role of galaxy mergers on quasars and galaxy evolution.

\section{Summary and Conclusion} \label{sec:concl}
We observed a $z=2.17$ dual quasar, \target, with ALMA Band 4 to detect the molecular gas and the underlying component. We detect the \colambda\ emission at 145.434 GHz and the nearby continuum at 143.538 GHz or at rest-frame 455 GHz. 

We find that the \coline\ emission has a kinematically and morphologically complex structure. Overall, the single-component \coline\ line fits indicates that the $v_{50}$ velocity map is similar to that of the ionized gas seen with \ha\ \citep{Ishikawa2024} with a red-/blue-shifted velocity gradient. From the detailed global line fits, we find that the CO gas can be separated into two distinct kinematic components: a narrow line ($w_{80}\sim150\ \textrm{km s}^{-1}$) component and a broad line ($w_{80}\sim320\ \textrm{km s}^{-1}$) component. 

The CO kinematics provides compelling evidence to mark a significant step forward in the lensing debate surrounding \target. Kinematic analyses contradict the predictions from the lensing hypothesis supporting the dual quasar interpretation the spectral dis-similarity of the CO emission (Figure \ref{fig:aperspect}) and an extended CO that is morphologically and kinematically complex that eliminates the lens-like CO ring (Figures \ref{fig:n1fits} and \ref{fig:n2fits}). The CO ring initially shown in the integrated intensity map is likely an artifact due to the blending of the kinematically distinct emission line components. Thus, we argue that \target\ is indeed a dual quasar. %Differentiating between a dual quasar and a lensed quasar requires significant multiwavelength observations. 

We examine the \vsigRat\ ratio of the narrow line and broad line gas components to assess the dynamical state of the CO gas. The \vsigRat\ maps suggest that the CO gas in \target\ is kinematically disturbed with two distinct dynamical phases: a dynamically complex, slow-moving gas and a turbulent, fast-moving gas. The CO kinematics suggest that \target\ is likely undergoing a galaxy merger event. However, some ambiguities remain in the exact kinematic interpretation of each component. This interpretation contradicts the gas rotation seen in the ionized gas phase \citep{Ishikawa2024}. It may be possible that the ionized and molecular gas phases trace different dynamics.  %Although the ALMA analysis 

We calculate a total molecular mass of $M_{H_2}=(6\pm5)\times10^{10}\ M_{\odot}$, in which the uncertainty is dominated by the different choice of conversion factors. If we compare the known SFR ($\sim1000\ M_{\odot}\textrm{ yr}^{-1}$) with \Mmol, we find that the ratio is above the Kennicutt-Schmidt relation for main-sequence galaxies \citep{Kennicutt1998}. This suggests that the host galaxy environment of \target\ is undergoing rapid star formation that exceeds the available fuel, possibly driven by the merger dynamics. Since no quasar-driven outflows have been detected, it is unlikely that the CO depletion is driven by quasar feedback. 

The rest-frame 455 GHz continuum is mostly dominated by the two compact cores that are co-spatial with the quasars. There is also an extended continuum component, stretching asymmetrically to the southwest by nearly 8 kpc. The \COcont\ ratio shows that the nuclear regions of the quasars are dominated by the continuum emission, whereas the \coline\ emission dominates in the outskirts. The CO/continuum ratio and the SFR/$M_{\textrm{H}_2}$ ratio (from \citealt{ChenYC2024}) indicate possible CO depletion. Interestingly, the nuclear 455 GHz continuum cannot be explained by an extreme starburst model or by the extrapolated synchrotron emission. Instead, the observed continuum may be associated with the excess emission from nuclear synchrotron radiation and possibly cold dust, whereas the outskirts are likely dominated by star formation. %\red{\target\ may be experiencing elevated quasar and star formation activity.}

This study, combined with previous JWST observations, demonstrates the power of spatially-resolved spectroscopy in significantly advancing our understanding of dual quasars. Although ambiguities in the kinematic interpretation remain, ALMA provides evidence of complex kinematics supporting the merger interpretation. Future observations of the stellar continuum and higher angular resolution may help address the ambiguity. The merger activity may fuel the intense starburst activity in \target\ and the two quasars. However, it is unclear whether this behavior is representative of the dual quasar population or if \target\ is an outlier. Detailed statistical studies is needed to address the mechanisms of dual quasars and the merger-quasar connection. 

\begin{acknowledgments}
The National Radio Astronomy Observatory is a facility of the National Science Foundation operated under cooperative agreement by Associated Universities, Inc. This paper makes use of the following ALMA data: ADS/JAO.ALMA\#2022.1.01427.S. ALMA is a partnership of ESO (representing its member states), NSF (USA) and NINS (Japan), together with NRC (Canada), MOST and ASIAA (Taiwan), and KASI (Republic of Korea), in cooperation with the Republic of Chile. The Joint ALMA Observatory is operated by ESO, AUI/NRAO and NAOJ.

Y.I. acknowledges support from the Space@Hopkins Graduate Fellowship. Y.I., N.L.Z., Y.-C.C., A.V., H.-C.H., and S.S. were supported in part by JWST-GO-02654 grant by the Space Telescope Science Institute. Y.S.~and X.L.~acknowledge support by NSF grant AST-2108162. Y.I. thanks D.~Coe, T.~Heckman, D.~Neufeld, C.~Norman, M.~Onoue, J.~Silverman, and M.~Yue for useful discussions.

\end{acknowledgments}


%\vspace{5mm}
\facilities{ALMA}


\software{\texttt{astropy} \citep{astropy2013,astropy2018,astropy2022}, \textit{CASA} \citep{casa2022}}

\bibliography{vodka_v2}{}

\bibliographystyle{aasjournal}

\end{document}